\documentclass[twocolumn]{article}
\pdfoutput=1
\usepackage{amsmath,amsfonts,slashed}
\usepackage{subcaption}
\usepackage{placeins}
\usepackage{graphicx}
\usepackage{booktabs}
\usepackage[dvipsnames]{xcolor}
\usepackage{mathrsfs} 
\usepackage{jheppub}
\usepackage{balance}

\usepackage{tikz}
\usetikzlibrary{trees}
\usetikzlibrary{decorations.pathmorphing}
\usetikzlibrary{decorations.markings}

  \definecolor{jblue}  {RGB}{20,50,100}
  \definecolor{lblue}  {RGB}{40,100,240}
  \definecolor{npurple}  {RGB} {153, 51, 204}
  \definecolor{wred}   {RGB}{217,0,56}
  \definecolor{white}   {RGB}{255,255,255}

  \definecolor{korange}   {RGB}{235, 80,  43}
  \definecolor{korange2}   {RGB}{245, 100,  63}
  \definecolor{kyelloworange}   {RGB}{255, 210,  110}
  \definecolor{kyelloworange2}   {RGB}{240, 170,  90}
  \definecolor{kred}   {RGB}{204,  102, 153}
  \definecolor{kpurple}   {RGB}{153,  61, 190}
  \definecolor{kpurplelight}   {RGB}{213,  161, 230}

  \tikzset{
      straightBit/.style={draw=npurple,very thick,postaction={decorate}, decoration={markings, mark=at position.50 with {\arrow[draw=npurple]{<}}}},
      straightBit2/.style={draw=npurple,very thick,postaction={decorate}, decoration={markings, mark=at position.60 with {\arrow[draw=npurple]{>}}}},
    photon/.style={decorate, decoration={snake}, draw=npurple,very thick},
    boson/.style={decorate, decoration={snake}, draw=npurple,very thick},
    electron/.style={draw=jblue,very thick, postaction={decorate},
             decoration={markings,mark=at position .55 with {\arrow[draw=jblue]{>}}}
    },
    electron2/.style={draw=jblue,very thick, postaction={decorate},
             decoration={markings,mark=at position .55 with {\arrow[draw=jblue]{<}}}
    },
    fermion/.style={draw=jblue,very thick, postaction={decorate},
              decoration={markings,mark=at position .55 with {\arrow[draw=jblue]{}}}
    },
    gluon/.style={decorate, draw=korange,very thick, 
      decoration={coil,amplitude=4pt, segment length=6pt}},
    higgs/.style={draw=wred,very thick, postaction={decorate},
             decoration={markings,mark=at position .55 with {\arrow[draw=wred]{>}}}
    },
    nothing/.style={draw=white,very thick}
  }

\graphicspath{{figures/}}
\notoc

\definecolor{magentaa}{rgb}{1,0,1}

\renewcommand{\vec}[1]{\mathbf{#1}}
\renewcommand{\vec}[1]{\boldsymbol{#1}}

\newcommand\hide[1]{}

\newcommand\intlumi{20~$\text{fb}^{-1}$}

\newcommand\ptmiss{\ensuremath{{\vec{\slashed{p}}}_T}}

\newcommand\lambdaptwothreeone{\ensuremath{\lambda'_{231}}}

\newcommand\GeV{{\mathrm{GeV}}}

\newcommand\ttbar{\ensuremath{t\bar{t}}}
\newcommand\wjets{\ensuremath{W+\text{jets}}}
\newcommand\wfake{\ensuremath{W+\text{fake}}}

\newcommand\epmm{\ensuremath{e^+\mu^-}}
\newcommand\emmp{\ensuremath{e^-\mu^+}}

\newcommand\madgraph{\textsc{MadGraph5\_aMC@NLO}}
\newcommand\madgraphshort{\textsc{MadGraph}}
\newcommand\madgraphver{2.4.3}
\newcommand\pythiaSix{\textsc{Pythia6}}
\newcommand\delphes{\textsc{Delphes}}
\newcommand\delphesver{3.3.3}
\newcommand{\summt}{\mbox{\ensuremath{m_T(e)+m_T(\mu)}}}
\begin{document}
\title{
    Difference between two species of emu hides a
    test for lepton flavour violation
}
\author{Christopher G.~Lester,}
\author{Benjamin H.~Brunt}
\affiliation{University of Cambridge, Department of Physics, Cavendish Laboratory, JJ Thomson Avenue, Cambridge, CB3 0HE, United Kingdom}
\abstract{
    We argue that an LHC measurement of some simple quantities related to the ratio of rates of $e^+ \mu^-$ to $e^- \mu^+$ events is surprisingly sensitive to as-yet unexcluded R-parity violating supersymmetric models with non-zero \lambdaptwothreeone\ couplings.  The search relies upon the approximate lepton universality in the Standard Model, the sign of the charge of the proton, and a collection of favourable detector biases.  The proposed search is unusual because:
   it does not require any of the displaced vertices, hadronic neutralino decay products, or squark/gluino production relied upon by existing LHC RPV searches;
    it could work in cases in which the only light sparticles were smuons and neutralinos;
    and
    it could make a discovery (though not necessarily with optimal significance) without requiring the computation of a leading-order Monte Carlo estimate of any background rate.
The LHC has shown no strong hints of post-Higgs physics and so precision Standard Model measurements are becoming ever more important. We argue that in this environment growing profits are to be made from searches that place detector biases and symmetries of the Standard Model at their core --- searches based around `controls' rather than around signals.
}
\maketitle

\section{Introduction}

The Large Hadron Collider (LHC) has not
yet seen any clear signs of physics beyond the Standard Model,
and has ruled out large parts of the parameter
spaces of models that were considered promising a decade ago.
Despite this slew of negative results,
is it possible that large signals could have remained hidden in plain sight? Are there any {\em simple} signatures that LHC collaborations have not yet checked?
The somewhat surprising answer to the latter question seems to be ``yes''.

We argue that overlooked and yet still interesting searches {\em can} be found
by using relatively simple detector-centred guiding principles.
We demonstrate the truth of this statement by following such a procedure concretely, and showing that it uncovers a simple (data only, model independent) but apparently overlooked lepton charge and flavour asymmetry search which is sensitive to departures from lepton flavour universality in the SM.   Post-facto, we show that the new search {\em happens} to be sensitive to a currently unconstrained part of RPV-supersymmetry parameter space.  Nonetheless, we regard the latter statement as being
of
only secondary importance
to the primary messages that (i) simple tests of SM symmetries are still  missing from the library of current results, and (ii) such tests can be found by exploiting, in a positive manner, sources of bias that at other times may seem to be confounding factors.

\newtheorem{mydef}{Definition}
\newtheorem{theorem}{Theorem}[section]
\newtheorem{corollary}{Corollary}[theorem]
\newtheorem{lemma}[theorem]{Lemma}

\section{Lepton charge-flavour symmetry at the LHC}

\subsection{Within the Standard Model}
\label{sec:firstThing}

Within the SM,
charged leptons (here $e^\pm$ and $\mu^\pm$ only) can be considered to be identical in all respects but mass.\footnote{In terms of the $W$-boson mass,  $m_e/m_W \approx 6.3\times 10^{-6} $ and $m_\mu/m_W \approx 1.3\times 10^{-4}$.}
However, no fundamental symmetry protects that universality or demands the absence of lepton flavour violation seen in the SM. Indeed, mixing in the neutrino sector (which is present in the SM) already violates it, though not at a level which is expected to be observable at the LHC. Given this, searching for signs of lepton flavour violation has long been considered a promising way to look for BSM physics.

Such searches cannot simply compare a distribution built with an electron requirement to an equivalent based on muons, since there are numerous places where {\em either} the $e$-$\mu$ mass difference, {\em or} some property of the detector or of the LHC itself, is expected to provide `boring' sources of flavour- or charge-dependent bias.  For example:
the ratio
$
\Gamma(\pi^+\rightarrow \mu^+ \nu_\mu)
/
\Gamma(\pi^+\rightarrow e^+ \nu_e)
\approx
8\times 10^5
$ and the greater penetrating power of muons in matter are both consequences of the $e$-$\mu$ mass difference.   That penetration asymmetry is also responsible for the existence of separate electron and muon detectors, and separate detectors can lead to differences between $e$ and $\mu$ acceptances, triggering rates, and reconstruction efficiencies.

Nonetheless, the {\em intrinsic} physics of the charged lepton sector in the SM is (so far as the LHC is concerned) CP-symmetric: for any flavour $l\in\{e,\mu\}$ large differences are not expected between the decay rates of $l^+$ and $l^-$, or between their production rates from neutral states.\footnote{Of course, {\em small} differences between positive and negative lepton production can be observed at the LHC as a result of CP-violation in the quark sector (e.g.~in neutral Kaon or $B$-meson mixing) but such observation requires
very carefully constructed analyses that are more complex than that we wish to propose here.}  This is not to say that LHC results are expected to be charge-symmetric. Many effects have a charge bias.  The proton-proton initial state has charge $+2$ leading to an excess of $W^+$ production over $W^-$ and so we expect to see more positive than negative leptons.  More subtle effects include: the small enhancement of positively charged cosmic ray muons at depth
(rock made of matter is better at shielding $\mu^-$ than $\mu^+$); the dominance of electrons over positrons in matter (e.g.~delta rays are always negatively charged); and the possibility that detectors themselves could sometimes have a greater acceptance or reconstruction efficiency for one charge over the other.\footnote{For example: in a muon detector with a similar design to that of ATLAS, a toroidal magnetic field would bend positive
and negative muons preferentially towards opposite ends of the detector.  In such a design, a $\mu^+$-$\mu^-$ reconstruction asymmetry could in principle arise if sensors at opposite ends of the detector had imperfectly matched efficiencies or acceptances.}

\subsubsection*{The strong LHC charge-flavour conspiracy}

But, hiding amid all these sources of charge and flavour bias lies a lucky charm, of sorts.  It is a dual consequence of the LHC beam and the SM itself.
This gift is the surprising fact that for any flavour-symmetric and suitably non-pathological event selection, every potentially significant bias or experimental uncertainty individually preserves the following property in the absence of other biases:
\begin{align}\label{eq:prop}
\frac{ \left<N(\mu^- e^+)\right> }{ \left<N(\mu^+ e^-)\right> }\le 1. \end{align}
We call this the `{\bf strong} LHC charge-flavour conspiracy'.
Note that the value `1' in the inequality above is the value that the ratio of expectations would take {\em if} there were no differences between electrons and muons.
\subsubsection*{The weak LHC charge-flavour conspiracy}
One can also define a `{\bf weak} 
LHC charge-flavour conspiracy'
by demanding that (\ref{eq:prop}) need only apply after {\em joint} rather than {\em individual} consideration of the same sources of bias and experimental uncertainty. 

\begin{lemma}It may be shown that the strong LHC charge-flavour conspiracy implies weak LHC charge-flavour conspiracy if every bias satisfies (\ref{eq:prop}) independently of the presence (or absence) of other biases.
\end{lemma}

Where does the strong conspiracy come from?
Some biases and experimental effects preserve the relationship (\ref{eq:prop}) by leaving the ratio of expectations invariant.  For example, if the reconstruction efficiency for electrons and positrons were independent of charge or any other property of the leptons in question,\footnote{We will come later to what happens when this assumption is invalid.} then any uncertainty in the reconstruction efficiency would change numerator and denominator by the same factor
leaving the ratio unchanged.
A second class of biases preserve (\ref{eq:prop}) by the simple expedient of making the ratio of expectations {\em smaller}.
For example: were it possible for delta-rays ($e^-$) to be detected as full tracks, this would increase the expectation in the denominator of the ratio only.
Finally, there is a third category of experimental effect or bias that can make the ratio {\em larger} rather than smaller, but by an amount that can be proved to be unable to take the ratio past unity.

To avoid interrupting the narrative here, we list in Appendix~\ref{sec:biases} the biases and experimental effects we have considered, together with arguments therein supporting the {\bf strong} conspiracy in each case.  Back in the body of the paper, however, our experimental method relies only on {\bf weak} conspiracy. Hereafter we therefore simply
take the {\bf weak} conspiracy to be a core
assumption and see where it leads.
\footnote{Note that the weak conspiracy is still likely to hold, even if some of the arguments in the Appendix turn out to be wrong, or if other sources of bias that do not satisfy strong conspiracy are found, provided that the `problematic' biases can be shown to be smaller than others for which the arguments remain valid.}

\begin{lemma}
It is trivial to show that the weak LHC charge-flavour conspiracy is equivalent to the statement ``$N(\mu^- e^+)\sim \text{Poiss}(\lambda_1)$ and $N(\mu^+ e^-)\sim \text{Poiss}(\lambda_2)$
    for some unknown parameters
    $ 0\le\lambda_1\le \lambda_2 < \infty$.''
   \label{lem:prop}
\end{lemma}


\subsection{Beyond the Standard Model}

\label{sec:rpv}

There is no reason that physics beyond the Standard Model need respect (\ref{eq:prop}).  For example, R-parity violating supersymmetric models may contain (among other things) `lambda prime' couplings.  An example of such a coupling is $\lambda'_{231}$ which introduces to the theory a vertex of the form
\begin{center}
\begin{tikzpicture}[scale=1,
                     thick,
             level/.style={level distance=3.15cm, line width=0.4mm},
             level 2/.style={sibling angle=60},
             level 3/.style={sibling angle=60},
             level 4/.style={level distance=1.4cm, sibling angle=60}
     ]
     \draw[jblue,very thick,electron] (-1.2,0) -- (0,0) ;
     \draw[jblue,very thick,electron] (0,0) -- (1,-1) ;
     \draw[jblue,very thick,dashed] (0,0) -- (1,+1) ;
\node[draw=none,fill=none] at (-0.6,-0.3){$d_R$}  ;
\node[draw=none,fill=none] at (0.2,+0.7){$\tilde \mu_L^-$}  ;
\node[draw=none,fill=none] at (0.2,-0.6){$t_L$}  ;
\end{tikzpicture}
\end{center}
\vspace{-2mm}
together with a similar vertex containing anti-particles rather than particles.\footnote{The $\lambda'_{231}$ coupling also introduces vertices containing a stop or a sbottom instead of a smuon.  We work, however, with a simplified model in which the only light sparticles are the left smuon and the neutralino, and so we neglect those other vertices.}  The best current limits \cite{Allanach:1999ic} on $\lambda'_{231}$ come from neutrino muon deep inelastic scattering data and demand
\begin{align}\lambda'_{231} < 0.18\times \frac{m_{\tilde b_L}}{100~\GeV}\end{align}
if the bottom squark is not decoupled.  In models where the bottom squark is not relevant, perturbativity can set other limits.  Requiring perturbativity at the weak scale forces $\lambda'_{231} <3.5$, while perturbativity all the way to the GUT scale leads to $\lambda'_{231} <1.5$.\footnote{Source: B.C.~Allanach, private communication, 2016.}  We choose to work in a simplified model where the only light sparticles are the smuon and the neutralino, so it is the last two limits that
are  most relevant to us.

When the neutralino is lighter
than the top quark, the presence of a non-zero $\lambda'_{231}$ coupling allows proton-proton collisions to produce muons in association with top quarks and missing transverse momentum\footnote{If the neutralino were heavier than the top quark, then the neutralino could itself decay to a muon, a top, and an anti-down quark by the reverse of the production process. This would eliminate missing transverse momentum from the signature, and introduce more leptons, and so is beyond the scope of this
paper.} (\ptmiss) via diagrams of the form:
\begin{center}
\begin{tikzpicture}[scale=1,
                     thick,
             level/.style={level distance=3.15cm, line width=0.4mm},
             level 2/.style={sibling angle=60},
             level 3/.style={sibling angle=60},
             level 4/.style={level distance=1.4cm, sibling angle=60}
     ]
     \draw[jblue,very thick,electron] (-1.2,0.5) -- (0,0.5) ;
     \draw[jblue,very thick,electron] (0,0.5) -- (0,-1) ;
     \draw[jblue,very thick,electron] (0,-1) -- (2,-1) ;
     \draw[jblue,very thick,dashed] (0,0.5) -- (1,0.5) ;
     \draw[jblue,very thick,electron] (1,0.5) -- (2,+1) ;
     \draw[jblue,very thick,photon] (1,0.5) -- (2,0) ;
     \draw[jblue,very thick,straightBit] (1,0.5) -- (2,0) ;
     \draw[jblue,very thick,gluon] (-1.2,-1) -- (0,-1) ;
\node[draw=none,fill=none] at (-1.4,0.5){$d$}  ;
\node[draw=none,fill=none] at (0.5,+0.9){$\tilde \mu_L^-$}  ;
\node[draw=none,fill=none] at (+0.34,-0.25){$t$}  ;
\node[draw=none,fill=none] at (-1.4,-1){$g$}  ;
\node[draw=none,fill=none] at (2.3,+1.1){$\mu^-$}  ;
\node[draw=none,fill=none] at (2.3,-0){$\chi^0_1$}  ;
\node[draw=none,fill=none] at (2.3,-1){$t$}  ;
\end{tikzpicture}
\end{center}
and\vspace{-5mm}
\begin{center}
\begin{tikzpicture}[scale=1,
                     thick,
             level/.style={level distance=3.15cm, line width=0.4mm},
             level 2/.style={sibling angle=60},
             level 3/.style={sibling angle=60},
             level 4/.style={level distance=1.4cm, sibling angle=60}
     ]
     \draw[jblue,very thick,electron] (0,0.5) -- (-1.2,0.5) ;
     \draw[jblue,very thick,electron] (0,-1) -- (0,0.5)  ;
     \draw[jblue,very thick,electron] (2,-1) -- (0,-1) ;
     \draw[jblue,very thick,dashed] (0,0.5) -- (1,0.5) ;
     \draw[jblue,very thick,electron] (2,+1) -- (1,0.5) ;
     \draw[jblue,very thick,straightBit2] (1,0.5) -- (2,0) ;
     \draw[jblue,very thick,photon] (1,0.5) -- (2,0) ;
     \draw[jblue,very thick,gluon] (-1.2,-1) -- (0,-1) ;
\node[draw=none,fill=none] at (-1.4,0.5){$\bar d$}  ;
\node[draw=none,fill=none] at (0.5,+0.9){$\tilde \mu_L^+$}  ;
\node[draw=none,fill=none] at (+0.34,-0.25){$t$}  ;
\node[draw=none,fill=none] at (-1.4,-1){$g$}  ;
\node[draw=none,fill=none] at (2.3,+1.1){$\mu^+$}  ;
\node[draw=none,fill=none] at (2.3,-0){$\chi^0_1$}  ;
\node[draw=none,fill=none] at (2.3,-1){$\bar t$}  ;
\end{tikzpicture}.
\end{center}
Since the proton's parton distribution function for the down quark is larger than that for the anti-down, the first diagram provides a larger contribution than the second, leading to more production of $\mu^-$ than $\mu^+$.  This is not by a factor only marginally more than one, but by an factor of order three to ten!\footnote{See Figure~\ref{fig:osdfleptonpair:mthists} later.}
The top or anti-top in the final state will also decay.  The top's final state will not always include leptons, but when it does they (i) will be of opposite sign to the smuon's muon, and (ii) will be equally likely to be electrons or muons. The effect of a non-zero value for $\lambda'_{231}$ is thus two-fold:  it both (a) increases the production of $\mu^- e^+$ by more than it increases the production of $\mu^+ e^-$, and (b) leads to an additional non-SM source of $\mu^- \mu^+$
events.  It is effect (a) in which the present paper is primarily interested.
\label{sec:secondThing}

\subsection{Existing constraints on such a model}
\label{sec:constraints}

\subsubsection*{Generic different flavour constraints: $\mu^\pm e^\mp$ }

So far as the authors are aware, there are no published LHC searches that make data-to-data comparisons of $\mu^- e^+$ and $\mu^+ e^-$ distributions of the form just proposed.
There are, however, many results published by LHC collaborations which relate to the {\em sum} (rather than difference) of those flavour combinations.

Variants include `opposite-sign different-flavour' (OSDF) searches and `no-charge-requirement different-flavour' (NCDF) searches.
OSDF examples include: the ATLAS RPV LFV $\lambda'_{312}$ search for a sneutrino resonance decaying to
an OSDF $\mu^\pm
e^\mp$ \cite{Aad:2011qr}; a later version of the same search that considers also $\lambda'_{321}$  \cite{Aaboud:2016hmk}; ATLAS searches for chargino and neutralino production \cite{Aad:2014vma}; a CMS dilepton invariant mass scan \cite{Khachatryan:2014fba}. NCDF examples include the CMS LFV Quantum Black Hole to $e\mu$ search \cite{Khachatryan:2016ovq}.
There is even an OSDF search from ATLAS \cite{Aad:2012yw} which targets LFV production caused by the simultaneous presence of two lambda prime couplings.  It requires $\lambda'_{131} \lambda'_{231}\ne 0$.

None of the above analyses is in direct competition with that proposed here as they collectively target absolute production rates, rather than differences.  Their sensitivity depends on many things, but is sometimes dominated by modelling uncertainties when Monte Carlo is used either for direct background prediction, or to extrapolate background rates from kinematically separate control regions.  These are very different methods to that proposed here.

\subsubsection*{Specific RPV-SUSY LFV constraints }

A recent review of LHC constraints on RPV couplings may be found in \cite{ATLAS-CONF-2015-018}. In relation to ${\bf L Q \bar D}$ couplings (its name for $\lambda'$ couplings) it notes that:\begin{quote}
    Searching for effects from ${\bf L Q \bar D}$  couplings, ATLAS has placed constraints on non-prompt decays leading to a multi-track displaced vertex \cite{Aad:2015rba}. A search for $\tilde t_1 \tilde t_1\rightarrow b l^+ \bar b l^-$ events also constrained prompt decays of the top squark via ${\bf L Q \bar D}$ couplings \cite{ATLAS-CONF-2015-015}. A similar model with non-prompt decays was investigated by CMS \cite{Khachatryan:2014mea}. The CMS search for events with multiple leptons and $b$-jets \cite{Chatrchyan:2013xsw} has been interpreted to constrain decays mediated by
    $\lambda'_{233}$ while Ref.~\cite{CMS-PAS-SUS-12-027} also examined $\lambda'_{231}$ decays.
    Furthermore the search in Ref.~\cite{Khachatryan:2014ura} constrained models with non-zero $\lambda'_{333}$ and $\lambda'_{3 j k}$ (with $j, k = 1, 2$), investigating signatures from $\tau$-leptons and $b$-jets.
\end{quote}
These studies either target displaced vertices (not a feature of our model), when one or more sparticles can travel a measurable distance before decaying, or target prompt decays of the neutralino.  For example, one study \cite{CMS-PAS-SUS-12-027} that set bounds on the same $\lambda'_{231}$ coupling we ourselves consider did so by looking for neutralino pair production followed by decays of the form:
\begin{center}
\begin{tikzpicture}[scale=1,
                     thick,
             level/.style={level distance=3.15cm, line width=0.4mm},
             level 2/.style={sibling angle=60},
             level 3/.style={sibling angle=60},
             level 4/.style={level distance=1.4cm, sibling angle=60}
     ]
     \draw[jblue,very thick,photon] (-1.2,0) -- (0,0) ;
     \draw[jblue,very thick,straightBit2] (-1.2,0) -- (0,0) ;
     \draw[jblue,very thick,electron] (0,0) -- (1,-0.6) ;
     \draw[jblue,very thick,electron2] (0,0) -- (1,0) ;
     \draw[jblue,very thick,electron] (0,0) -- (1,+0.6) ;
\node[draw=none,fill=none] at (-0.6,-0.3){$\chi^0_1$}  ;
\node[draw=none,fill=none] at (1.3,+0.7){$\mu^-$}  ;
\node[draw=none,fill=none] at (1.2,0.0){$\bar d$}  ;
\node[draw=none,fill=none] at (1.2,-0.6){$t$}  ;
\end{tikzpicture}
\end{center}
which are not present in our model due to our neutralinos being lighter than the top quark.
There are therefore no existing searches that claim to be sensitive to the $\lambda'_{231}$ coupling in a model of the sort we have discussed.

\subsubsection*{Other constraints}

Though our model contains a non-zero R-parity-violating $\lambda'_{231}$ coupling, it still has all the other parts of the model which respect R-parity.  In principle, therefore, the model we have proposed is constrained by all searches that have considered di-slepton production and decay to neutralinos in the context of simplified models.  For example, both ATLAS \cite{Aad:2014vma} and CMS \cite{Khachatryan:2014qwa} have ruled out some of the left-slepton masses below 300~GeV under the assumption that the smuon and selectron are mass-degenerate.
Our proposal has sensitivity to much higher left-slepton masses (perhaps even up to 2~TeV) and so is complementary to those existing searches, though of course it relies upon the RPV sector to accomplish that extension.

Additionally, by the so-called `effect (b)' mentioned at the end of Section~\ref{sec:secondThing}, our model predicts an overall increase in $\mu^\pm\mu^\mp$ production not matched by any increase in $e^\pm e^\mp$. That excess is potentially observable by any of the LHC analyses that have looked at di-muon spectra, including \cite{
Aad:2014cka,ATLAS-CONF-2016-045,Khachatryan:2014fba,Khachatryan:2016zqb,Chatrchyan:2012oaa,
Aad:2014wca,Aaboud:2016qeg,Aad:2014vma,CMS:2014hka,Chatrchyan:2012hda}, and in particular
those which rely on additional handles such as \ptmiss, $M_{T2}$, or $H_T$, given the non-resonant nature of our signal.  Although it might be interesting to see whether any of those searches have sensitivity to our model, the aim of this paper to motivate interest in charge-flavour asymmetry searches, not to determine how best to discover non-zero $\lambda'_{231}$.  We therefore leave this question unanswered, noting that the answer would be in any case be irrelevant for BSM models that produce
a charge-flavour asymmetry without a
flavour asymmetry.\footnote{Charged Higgs bosons might decay at different rates to each of $e\nu_e$, $\mu\nu_\mu$ and $\tau\nu_\tau$.  Accurate measurements of cross section ratios such as $\sigma(e\tau)/\sigma(e\mu)$ or $\sigma(e\mu)/\sigma(\mu\tau)$ are therefore sensitive to charged Higgs production \cite{Aad:2012rjx,Palmer:2013hgn}. Though such searches share some features with ours (principally an interest in different flavour final states) they are posed as ratio measurements where the only
difference between numerator and denominator is flavour, not charge-flavour. These analyses do not therefore tread on our toes either.}



\subsection{Summarising remarks}



\begin{itemize}
        \item
            In Section~\ref{sec:firstThing} we saw  that the Standard Model makes $\mu^+ e^-$ and $\mu^- e^+$ events at very similar rates but has a (potentially very small) bias toward one charge combination.
        \item
In Section~\ref{sec:secondThing} we saw that at least one straw BSM theory (presumably there are many
more) can favour the {\em other} charge combination, and by a much larger factor than the SM.
\item In Section~\ref{sec:constraints} we saw that the straw BSM theory contained features that are untouched by existing searches.
\end{itemize}

The above remarks tell us that comparisons between $\mu^+ e^-$  and $\mu^- e^+$ distributions are not dull. They can contain readily accessible information about the lepton-flavour symmetry that is unexploited at present.

\newcommand\freegift{\footnote{It is, of course, possible that requests to compare $\mu^- e^+$ and $\mu^+ e^-$ rates {\em have} been made in the theory literature, but have failed to generate action within LHC collaborations and have also evaded the authors' attempts to find them.  If that is the case, the authors are prepared to wager that any such paper has not also pointed out the utility of the expected charge bias described in Section~\ref{sec:firstThing}.  Persons believing the authors
to be mistaken are encouraged to let them know. The first supplier of a reference to a paper providing a counter example to the statement of the wager shall, if that paper was published in a peer-reviewed journal before 1st December 2016, be entitled to receive, at the authors' expense, a four-course dinner and one night's accommodation in a Cambridge college upon his or her next visit to the UK. Terms and conditions apply.}}

It appears that, for unknown reasons, this
search strategy has either received no attention, or at the very least has received less attention than it deserves. This seems surprising, given the simplicity of the suggested comparison and the status of lepton-flavour conservation as an {\em unprotected} symmetry of the Standard Model. Perhaps this underlines the nature of the remarks made in the introduction about the need to make the tests that are motivated by `detector-centred guiding
principles' and `fundamental symmetries' as these are, at present, few and far between.\freegift

\label{sec:three}

\section{Illustration of viability }

There are many different ways that LHC experiments could choose create analyses based on charge-flavour $e^\pm \mu^\mp$ asymmetries.  Some might prioritise reach for a particular LFV model. Some might prefer to measure only the intrinsic SM asymmetry.  Others might prefer robustness and simplicity of analysis design over discovery reach in a particular model.  Each BSM model motivates a different search variable (\ptmiss, $M_T$, etc.) in which to look for the asymmetry.  Each LHC experiment has
its own particular idiosyncrasies, detector-induced asymmetries and sources of systematic uncertainty which would need full consideration by methods specific to itself.  Lastly, each experiment would rightly want to get the most out of any data by using the best available statistical techniques at its disposal.

While very important, all those choices and issues are beyond the scope of this paper.  The main intention of this paper is to raise interest and awareness in the benefits of using charge-flavour $e^\pm \mu^\mp$ asymmetries to find BSM effects.  Accordingly, we choose here to illustrate the viability of the general proposal with the {\em simplest} statistical methods available to us, even though the methods used by real experiments would assuredly be considerably more developed.
Though our illustration focuses narrowly on exclusion of the SM using a hypothesis test motivated by a particular class of LFV model, real usage will be different!

\subsection{Selected scope}

\begin{figure}
	\begin{center}
		\begin{subfigure}{.48\textwidth}
			\includegraphics[width=\linewidth]{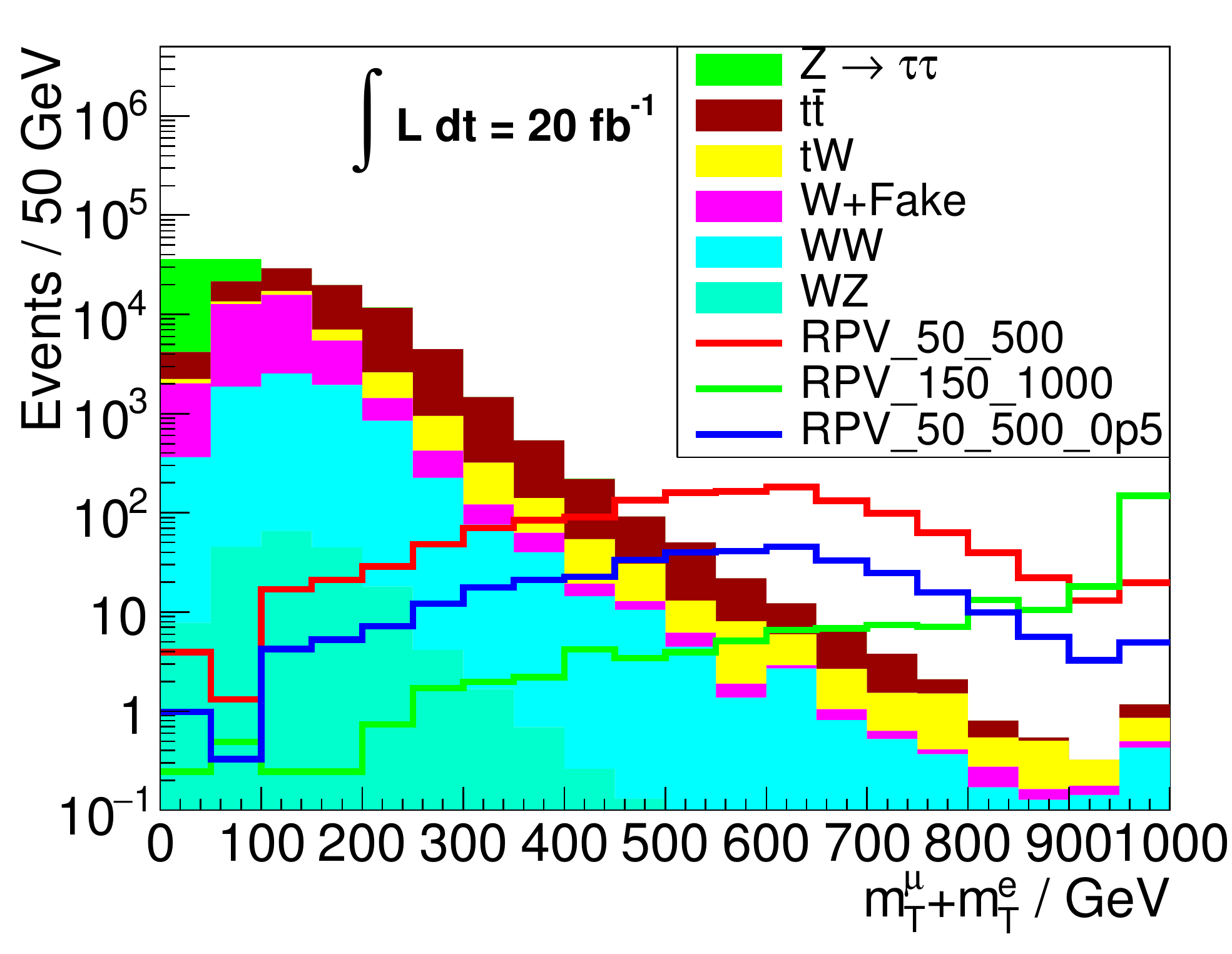}
			\caption{\summt\ in \epmm\ events}
			\label{fig:osdfleptonpair:mthists:mumelp_mt}
		\end{subfigure}
		\begin{subfigure}{.48\textwidth}
			\includegraphics[width=\linewidth]{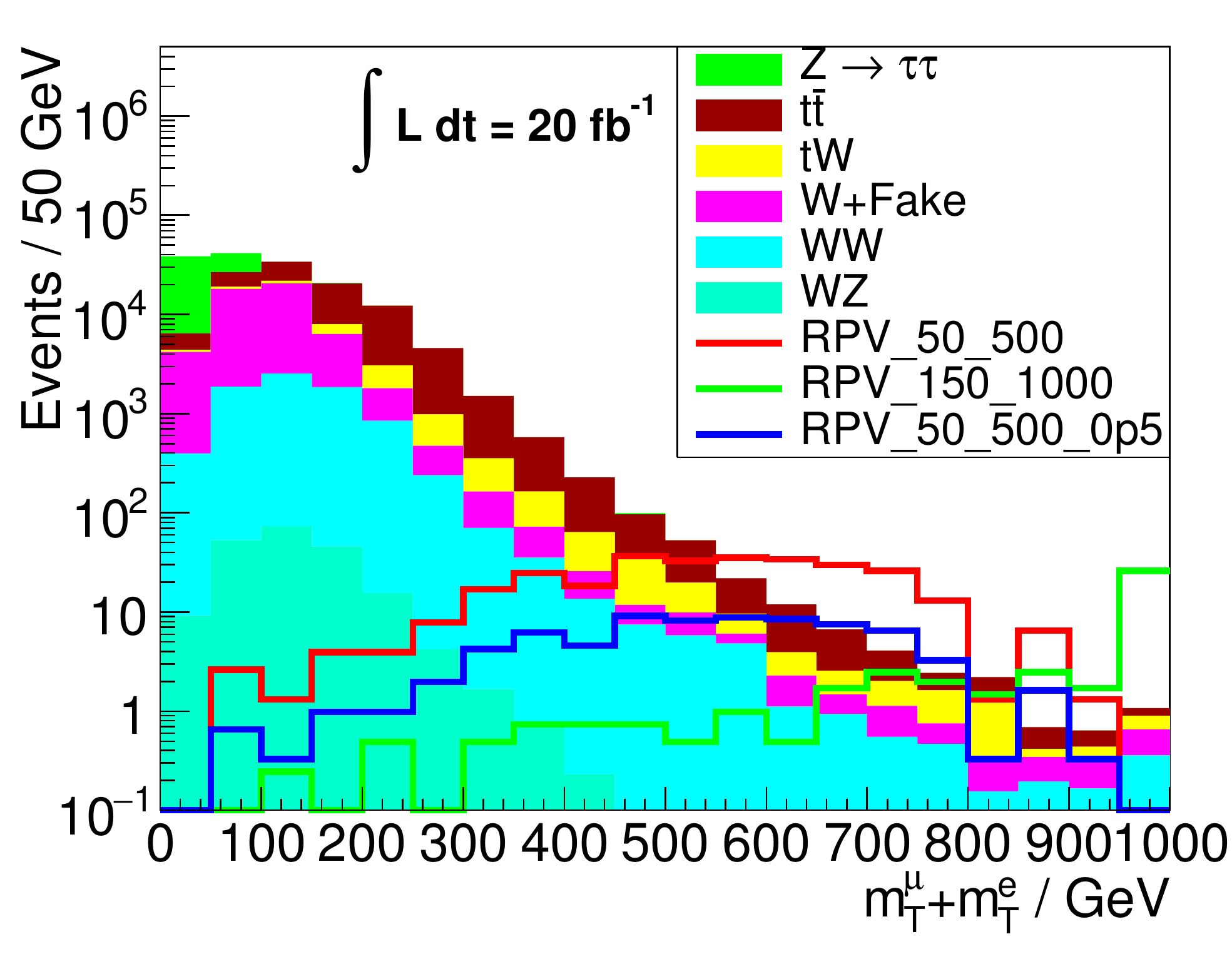}
			\caption{\summt\ in \emmp\ events}
			\label{fig:osdfleptonpair:mthists:mupelm_mt}
		\end{subfigure}
		\caption{The expected distributions of \summt\ in events with OSDF leptons (\epmm\ and \emmp\ in (\subref{fig:osdfleptonpair:mthists:mumelp_mt}) and (\subref{fig:osdfleptonpair:mthists:mupelm_mt}) respectively).}
		\label{fig:osdfleptonpair:mthists}
	\end{center}
\end{figure}

\begin{figure}
	\begin{center}
		\includegraphics[width=\linewidth]{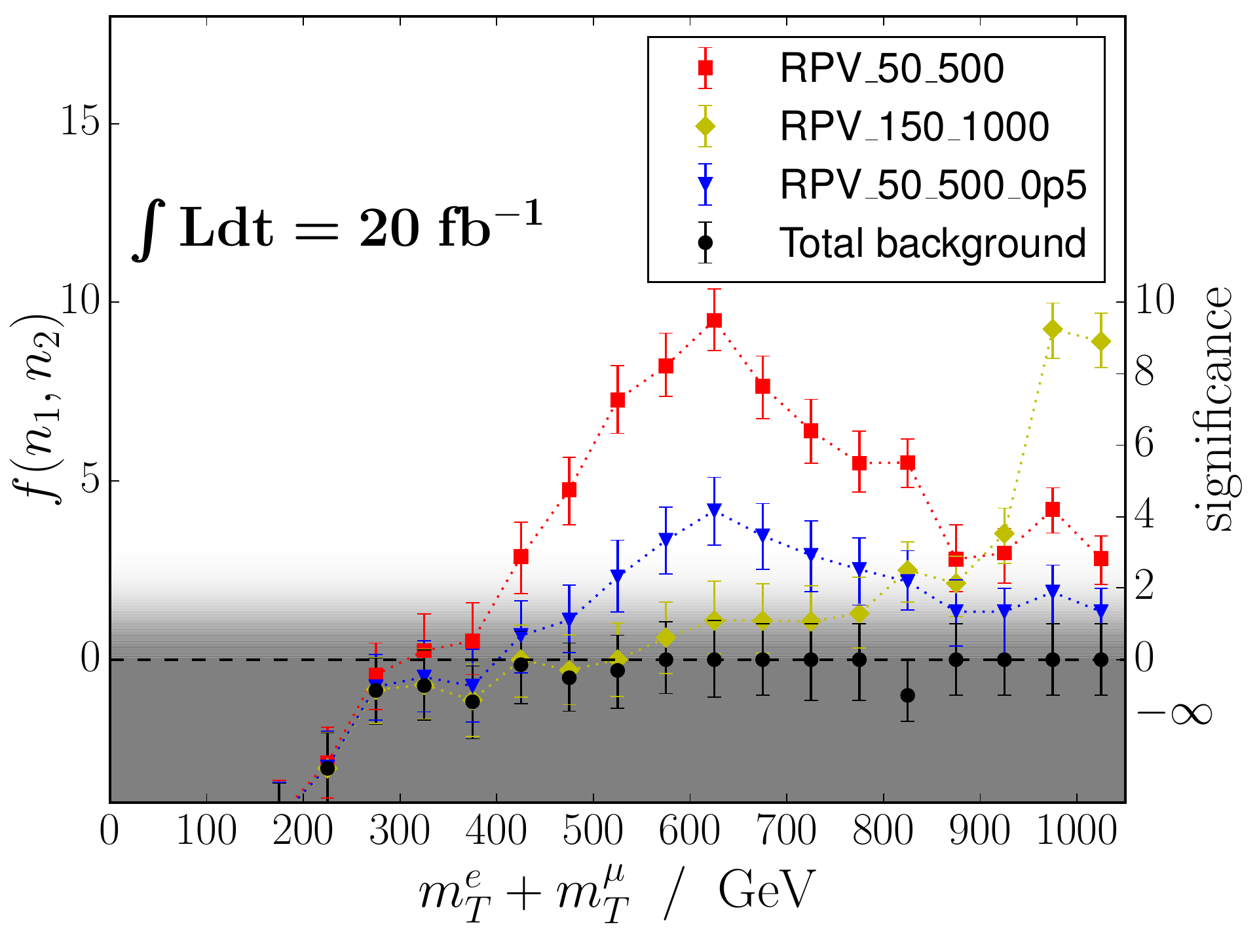}
        \caption{The left-hand axis shows the median value of the statistic $f(n_1,n_2)$ in each 50~GeV bin of \summt, while the right-hand axis shows the mapping of these values to `sigmas significance' using the blue line of Figure~\ref{fig:pvalue-numerical-limiting}.
			The black points show background alone, and the coloured points show the sum of the background with each of the example signals.
            Error bars indicate the $50\pm34$th percentile values of $f$, i.e.~the $\pm1 \sigma$ deviations from the median.
			The shaded region indicates the null hypothesis of $f(n_1,n_2) \le 0$ and unit variance upwards.
			The dotted lines connecting points are given as a guide to the eye.
		}
		\label{fig:osdfleptonpair:mumelp_mupelm_sig_mt}
	\end{center}
\end{figure}

\label{sec:osdfleptonpair}

We illustrate the likely utility of an $e^\pm\mu^\mp$ charge-flavour asymmetry in the context of models having the R-parity-violating (RPV) coupling $\lambda'_{231}$ described earlier.  The proposed search uses a selection targeting opposite-sign, different-flavour di-lepton events having large sum of transverse masses, \summt.  Each transverse mass in this sum is defined by $m_T(l) = \sqrt{2 \ptmiss p_T^l(1-\cos\theta)}$ wherein $\theta$ is the angle in the transverse plane between
the $\ptmiss$ and the lepton (or anti-lepton) $l$ concerned. Our signal model only produces large muonic transverse mass $m_T(\mu)$, however it is necessary to keep both electron and muon transverse masses in the sum to ensure that all signal regions remain flavour symmetric, as required by the charge-flavour conspiracy definition given earlier.

\subsection{Monte Carlo simulations}
\label{sec:montecarlo}

When studying the {\em expected} performance of the proposed method
it is necessary to use Monte Carlo simulations, even though the proposed method could avoid use of Monte Carlo when run on data. 
All the Monte Carlo samples produced for this purpose used \madgraph~\cite{madgraph}, version \madgraphver.
The generated samples were hadronised using \pythiaSix~\cite{Sjostrand:2006za} through the \textsc{Pythia-PGS} interface, with detector simulation provided by \delphes~\delphesver~\cite{delphes}.

\subsubsection{Signal simulation}
Models of RPV SUSY are simulated using \madgraphshort\ with additional model ``RPV MSSM''~\cite{rpvmssmUfo}.
All RPV couplings were set to zero except the coupling of interest, which, except where stated otherwise, was set to unity.\footnote{The cross section for the two two-to-three processes shown in Section~\ref{sec:secondThing} scales as the square of $\lambda'_{231}$.}
The masses of the left-handed smuon and the lightest neutralino were varied between models, while the masses of all other sparticles were set to large values, beyond the reach of the LHC, in order to decouple them.

The sensitivity studies in the following sections use a set of signal samples which form a ``grid'' in the plane of smuon mass and neutralino mass.
Neutralino masses above the top quark mass were not considered, to keep the neutralino stable on detector timescales.
Three models, with parameters shown in Table~\ref{tab:examplesignals}, are chosen as examples to be shown in figures.
\begin{table}
	\begin{tabular}{l c c c}
		\toprule
        Label on plots & ($m_{\tilde\mu}, m_{\tilde\chi^0_1})$ &  $\lambda'_{231}$ & $\sigma_{\text{RPV}}$  \\
                 & GeV &   & pb  \\
        \midrule
        RPV\_50\_500      & (\phantom{1}500, \phantom{1}50)  & 1.0 & 1.3\phantom{8}    \\
        RPV\_150\_1000    & (1000, 150) & 1.0 & 0.25  \\
        RPV\_50\_500\_0p5 & (\phantom{1}500, \phantom{1}50) & 0.5 & 0.33    \\
		\bottomrule
	\end{tabular}
    \caption{The example RPV SUSY models used in this document.
		}
	\label{tab:examplesignals}
\end{table}

\subsubsection{Background simulation}
There are several standard model processes which produce final states similar to the models of RPV SUSY.
The dominant Standard Model background comes from the production of top-quark pairs (\ttbar).
Also included is the Standard Model production of a top quark in association with a $W$ boson ($tW$), $Z/\gamma\rightarrow\tau\tau$ and diboson ($WW$ and $WZ$).

An additional background comes from single-lepton processes in which an additional lepton is gained by misidentification of a jet or similar mechanisms. We attempt to model the more prevalent process producing a ``fake'' electron using a sample of simulated $W$+jets events in which the $W$ produces a muon, and a jet is treated as an electron. The chance for this misidentification to occur is taken as 0.5\%, which is similar to the rate reported by the ATLAS collaboration in Ref.~\cite{ATLAS-CONF-2016-024}, assumed to be independent of the charge of the electron produced.

\hide{Section~\ref{sec:singlelepton} examines final states requiring a single lepton (an electron or a muon).
For this, the dominant background contribution comes from Standard Model production of $W$~bosons, which is simulated in slices of \ptmiss.
Top-pair and $tW$ processes are also included. \textcolor{Red}{$Z$ as well, since no veto on second lepton?}}

In each background process, samples were generated with zero and with one or more extra hard-process partons in the final state.
The \textsc{MadEvent} matrix element was matched to the parton shower using the shower-$k_T$ scheme~\cite{showerkt} with $p_T$-ordered showers.
The matching scale was set to 80~GeV for the \ttbar\ and $tW$ processes, and to 30~GeV for the $W$, $Z$ and diboson processes.



\subsection{Illustrative analytic framework}

\label{sec:stats}

Given the caveats mentioned at the start of Section~\ref{sec:three}, we elect to illustrate the viability of the search using a hypothesis test that seeks to accept or reject the SM.  The test uses the two Poisson random variables $N(\mu^- e^+)$ and $N(\mu^+ e^-)$ in Lemma~\ref{lem:prop} which (after an appropriate selection) we abbreviate as $N_1$ and $N_2$ respectively.  The null hypothesis $H_0$ of our test is, in effect, the statement  $0\le\lambda_1\le\lambda_2<\infty$, concerning the means of $N_1$ and
$N_2$, while the alternative hypothesis
$H_1$ is that $0<\lambda_1>\lambda_2\ge0$.  Within this paper, $H_1$ is used only to the extent that it motivates the choice of the test statistic.  The quantitative results we report concern only the probabilities of fluctuations under $H_0$ (the background hypothesis) of sufficient size to account for straw BSM models we simulate.\footnote{Any real experiment performing a charge-flavour $e^\pm \mu^\mp$ asymmetry measurement would probably use something closer to a likelihood ratio
    $p(N_1,N_2|H_1)/p(N_1,N_2|H_0)$, with  profiling over $\lambda_1$ and $\lambda_2$ in the appropriate places. Use of approach will inevitably lead to different sensitivities that we show herein, particularly
at the borders of sensitivity.  While such differences differences will be important for a real analysis, they are not important for our purposes of illustrating that the proposed searches are worth performing and have considerable sensitivity to some models.}
A `test statistic' is still needed by our illustrative test.  It must be a function $f(N_1,N_2)$.
Without loss of generality, we need only consider test statistics $f(N_1,N_2)$ for which {\em larger} values are increasingly suggestive of new physics. Given observed values $n_1$ and $n_2$ for random variables $N_1$ and $N_2$,
we therefore define our $p$-value under the null hypothesis, $p_0(f(n_1,n_2))$, as:
\begin{align}
    p_0 = \max_{0\le\lambda_1\le\lambda_2} P({f(N_1,N_2)\ge f(n_1,n_2)} \mid S(\lambda_1,\lambda_2)) \label{eq:pvalue}
\end{align}
where $S(\lambda_1,\lambda_2)$ is the statement that $N_1\sim \text{Poiss}(\lambda_1)$ and $N_2\sim \text{Poiss}(\lambda_2)$.\footnote{The `$\max$' in (\ref{eq:pvalue}) is necessary since the null hypothesis does not specify particular values for $\lambda_1$ and $\lambda_2$, only their relative size. Accordingly, all allowed values of $\lambda_1$ and $\lambda_2$ must be tested, and the least significant $p$-value reported.} So-defined, $p_0$ is the
probability that a {\em more extreme} value of the test statistic than that observed could have appeared under the most conservative interpretation of the null hypothesis (i.e. of the SM).

\begin{center}
	\begin{figure}[tbp]
		\includegraphics[width=\linewidth]{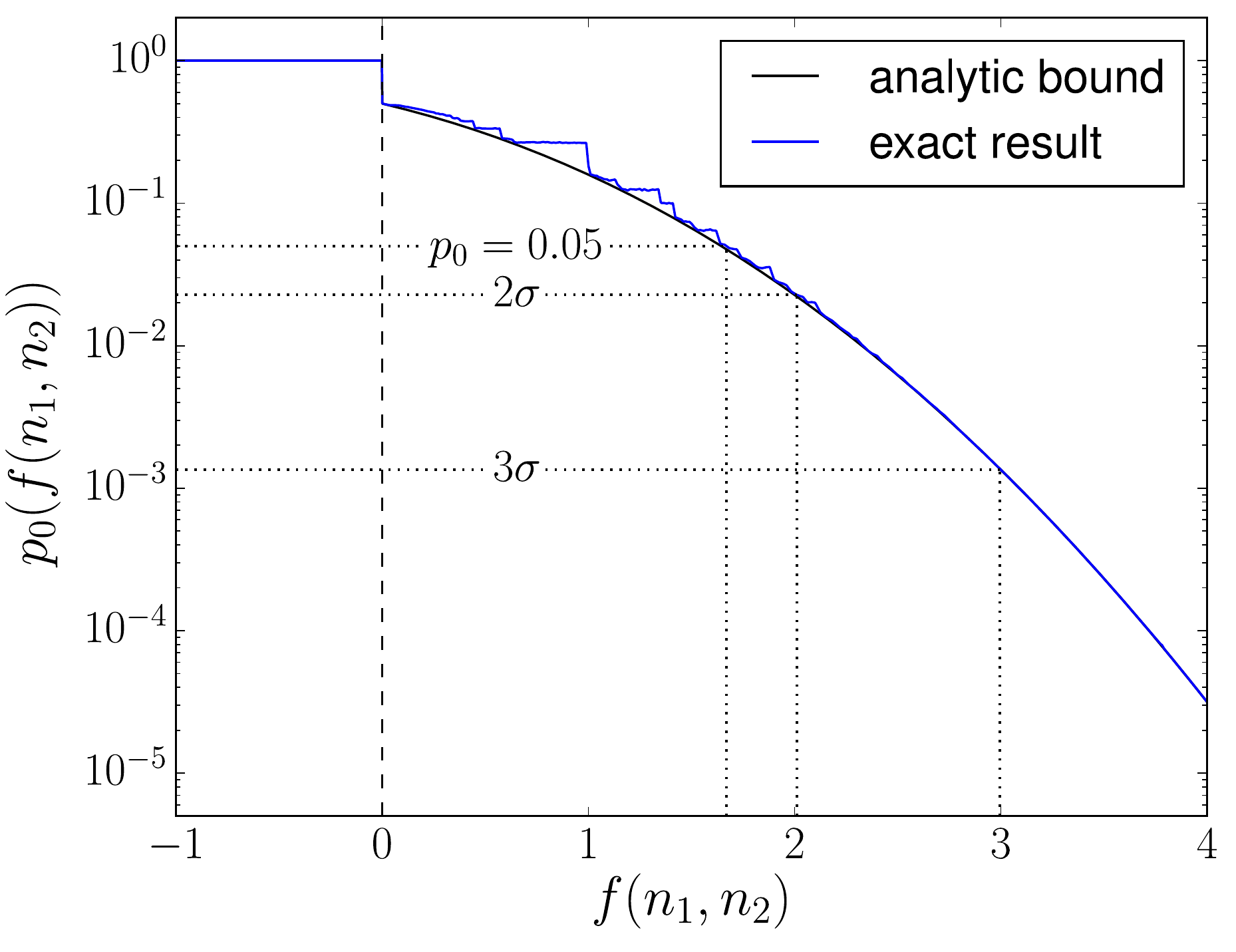}
		\caption{The $p$-value, defined in (\ref{eq:pvalue}), plotted as a function of the statistic $f(n_1,n_2)$.
		Also shown is the lower bound (\ref{eq:erf}) to which the $p$-value converges for large values of the statistic.
		}
		\label{fig:pvalue-numerical-limiting}
	\end{figure}
\end{center}

What function $f(N_1,N_2)$ should be used to define the test statistic?  
There is a large literature concerning hypothesis tests related to comparisons of Poisson means, some of which may be found in \cite{Krishnamoorthy200423}.  It is not the wish of this paper to get mired in questions of statistical optimality, however.  In any case, the alternative hypothesis $H_1$ must play an important role in selecting test statistics.  Without any claims to optimality, and supported by little more than  (i) self-evident differences between 
$H_0$ and $H_1$, and (ii) the desire to keep our illustration simple, 
we elect to use the test statistic:
\begin{align}
    f(N_1,N_2)=\begin{cases}
        \frac{N_1-N_2}{\sqrt{N_1+N_2}} & \text{if $N_1+N_2\ne 0$} \\
        0 & \text{otherwise.}
    \end{cases}
\end{align}
In the limit of large
$\lambda_1+\lambda_2$ the above choice becomes Gaussian distributed with mean $\lambda_1-\lambda_2$ and unit variance.\footnote{Note that for {\em any} value of $\lambda_1+\lambda_2>0$ the random variable $f_3(N_1,N_2)=\frac{N_1-N_2}{\sqrt{\lambda_1+\lambda_2}}$ has, by construction, unit variance and mean $\lambda_1-\lambda_2$.}
This property ensures that the statistic has well defined behaviour  under the infinite part of the maximisation performed in
(\ref{eq:pvalue}).
For this choice of $f$ it may be proved that
$p_0(f(n_1,n_2)) = 1$ if $n_1\le n_2$.
For $n_1>n_2$ it is necessary to evaluate $p_0(f(n_1,n_2))$ numerically. The resulting distribution is shown in Figure~\ref{fig:pvalue-numerical-limiting}, in which the bound
\begin{align}
    p_0(f(n_1,n_2)) \ge \frac 1 {\sqrt{2\pi}} \int_{f(n_1,n_2)}^\infty
    \!\!\!\!\!\!
    \!\!\!\!\!\!
    e^{-x^2/2} \text{d}x \label{eq:erf}
\end{align}
may be easily seen.\footnote{This bound stems from consideration of large $\lambda$  values in  (\ref{eq:pvalue}).} Note that the bounding function in (\ref{eq:erf}) is also a very good approximation to $p_0(f(n_1,n_2))$ when $f(n_1,n_2)\gtrsim 2.5$.
The local conclusion of this section is that, {\em if the weak conspiracy is valid}, it is possible to perform a hypothesis test of the apparent lepton flavour symmetry in the Standard Model.  That test requires one to count the numbers $n_1$ and $n_2$ of events in (respectively) $\mu^- e^+$ and $\mu^+e^-$ subsets of any common selection.  The null (SM) hypothesis may then be rejected if the value of $f=(n_1-n_2)/\sqrt{n_1+n_2}$ is smaller than any desired $p$-value, using the
translation curve shown in Figure~\ref{fig:pvalue-numerical-limiting}.  We note in passing that for any positive value of $f$ for which the black and blue curves in Figure~\ref{fig:pvalue-numerical-limiting} touch, $p_0$ is the probability that a normally-distributed 
random variable exceeds $f$; in the loose language used in experimental particle physics, $f$ counts `sigmas' of significance.

\subsection{Results}
\label{sec:results}

\begin{figure}
	\begin{center}
		\begin{subfigure}{.43\textwidth}
			\includegraphics[width=\textwidth]{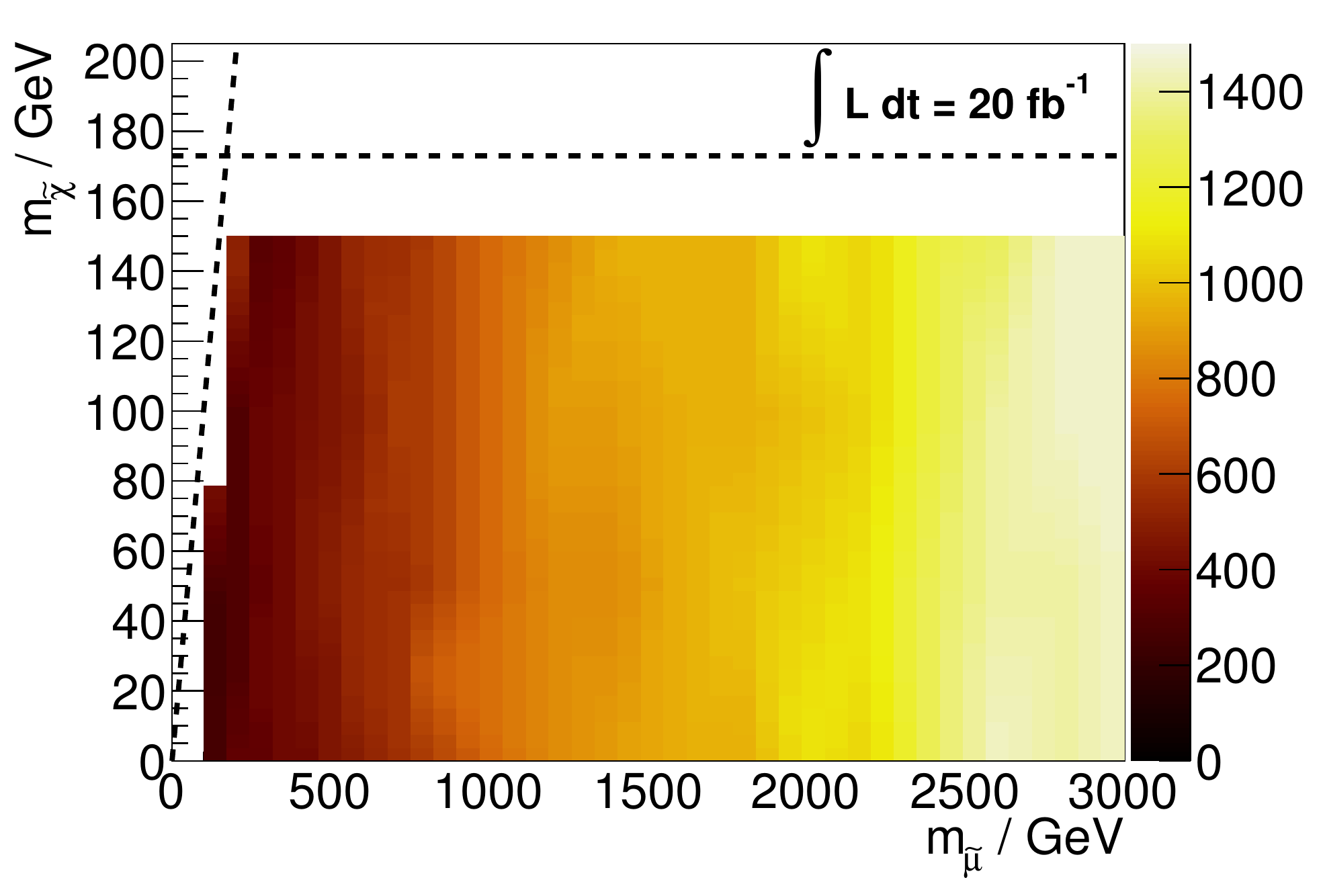}
		\end{subfigure}
		\caption{The \summt\ threshold (GeV) used to define the signal region for each of the $\lambda'_{231}=1$ signal model considered, set for each model so as to maximise the median sensitivity.
		}
		\label{fig:osdfleptonpair:mtcut}
	\end{center}
\end{figure}

\begin{figure}
	\begin{center}
		\begin{subfigure}{.43\textwidth}
			\includegraphics[width=\textwidth]{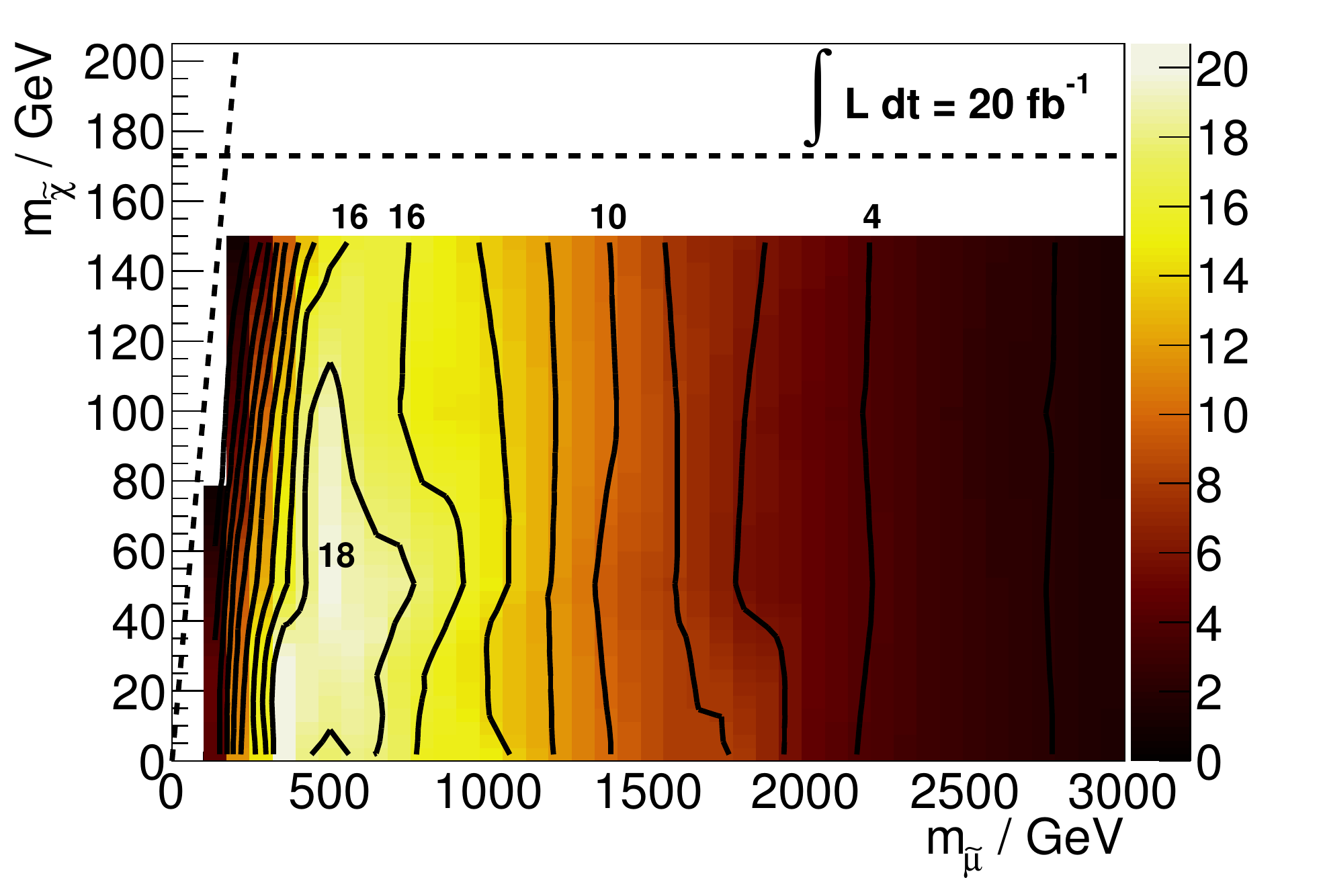}
		\end{subfigure}
		\caption{The median value of $f$ (in effect the expected sensitivity of the method) for the grid of $\lambda'_{231}=1$ signal models.  Recall that this $f$ statistic (unlike the bin-wise $f$ statistics of Figure~\ref{fig:osdfleptonpair:mumelp_mupelm_sig_mt}) aggregates all events with \summt\ greater than the model-dependent threshold shown in Figure~\ref{fig:osdfleptonpair:mtcut}.
			Contour lines show integer values of sensitivity.
		}
		\label{fig:osdfleptonpair:sensitivity}
	\end{center}
\end{figure}

\begin{figure}[tbp]
	\begin{center}
		\begin{subfigure}{.43\textwidth}
			\includegraphics[width=\textwidth]{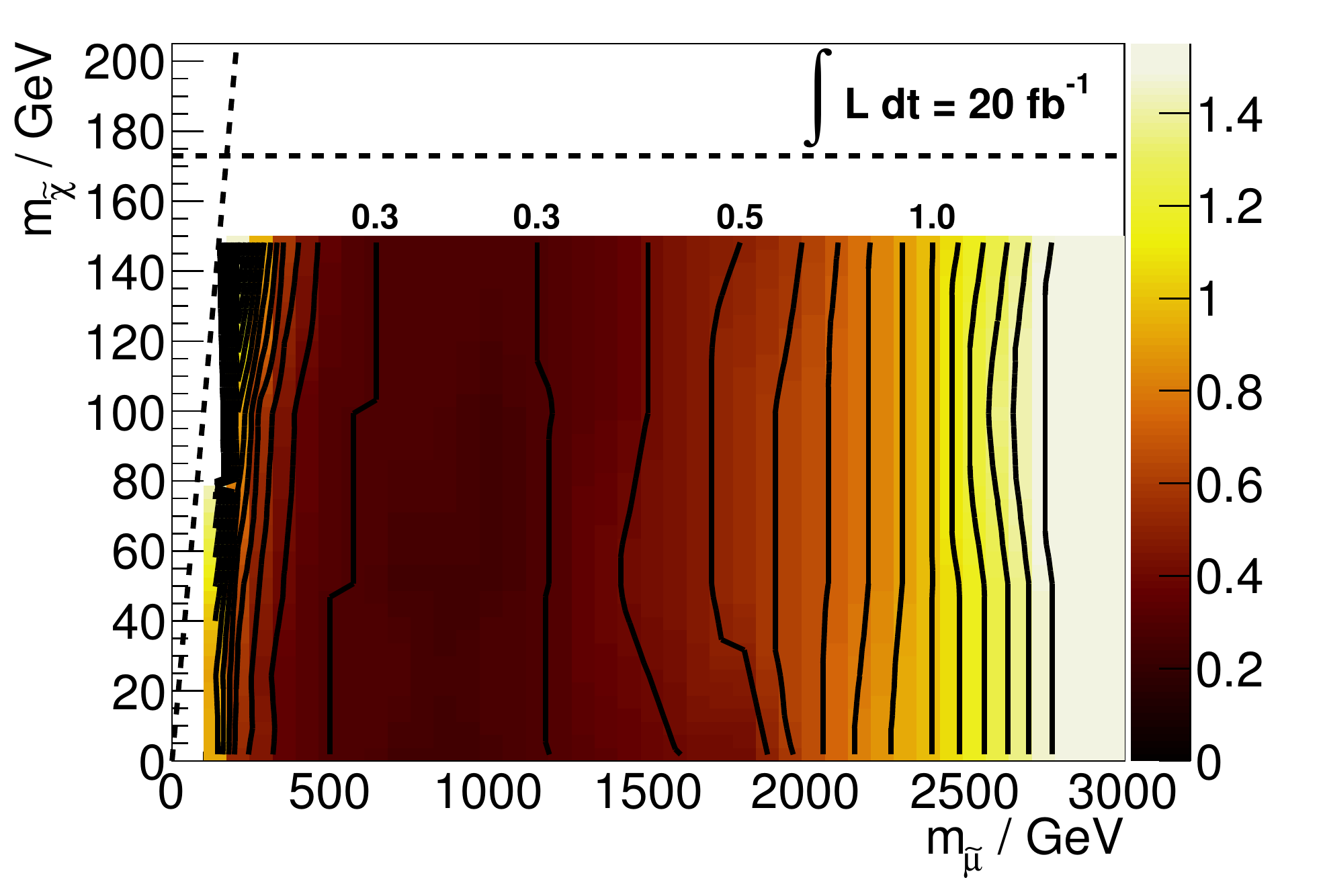}
		\end{subfigure}
		\caption{The minimal value in each model of the $\lambda'_{231}$ coupling for which a sensitivity of $f \ge 3$ is achieved.
			Contour lines are drawn at intervals of 0.1.
		}
		\label{fig:osdfleptonpair:couplingsensitivity}
	\end{center}
\end{figure}

Figure~\ref{fig:osdfleptonpair:mthists} shows the expected distributions of the transverse mass sum (\summt) for three example signal processes, together with the main SM backgrounds, for an effective integrated luminosity of \intlumi.
Comparing events with a negatively-charged muon (Figure~\ref{fig:osdfleptonpair:mthists:mumelp_mt}) and those with a positively-charged muon (Figure~\ref{fig:osdfleptonpair:mthists:mupelm_mt}), it can be seen that the signal models favour the production of $\mu^- e^+$ over $\mu^+ e^-$ by a factor of about three in the model with a lightest smuon, and by a factor of more than ten in the model with the heaviest smuon.
Figure~\ref{fig:osdfleptonpair:mumelp_mupelm_sig_mt} plots the median value of the statistic $f(n_1,n_2)$ for {\em each} 50~GeV bin independently of the others.
The medians here are taken over numerous draws (pseudo experiments) in each bin, assuming the event counts to be Poisson-distributed with mean set by the Monte Carlo predictions of Figure~\ref{fig:osdfleptonpair:mthists}.
Figure~\ref{fig:osdfleptonpair:mumelp_mupelm_sig_mt} demonstrates that by selecting events with sufficient \summt\ (and so suppressing the SM background), sensitivity to the charge asymmetry in the signal can be obtained in many bins.

Clearly the best sensitivity to any one model will be obtained not by using any one bin, but by using a set of them. For convenience we elect to use a regions of \summt\ starting at some threshold value and extending upwards to infinity.  For each member of the family of models living on the grid of smuon and neutralino mass values described earlier, we therefore determine a value for an \summt\ threshold that approximately optimises the median sensitivity for that model.
The \summt\ thresholds found are illustrated in Figure~\ref{fig:osdfleptonpair:mtcut}.
For models with a fixed value of the coupling $\lambda'_{231}=1$, the resulting sensitivity is shown in Figure~\ref{fig:osdfleptonpair:sensitivity}.
It shows: (i) that 3$\sigma$ median sensitivity to models with $\lambda'_{231}=1$ is expected for slepton masses between 350~GeV and about 2.5~TeV, and (ii) that the median sensitivity may be greater than 10$\sigma$ for slepton masses in the range 400--1400~GeV. As an alternative way of representing the same data Figure~\ref{fig:osdfleptonpair:couplingsensitivity} plots, for each model, the minimal value of the $\lambda'_{231}$ coupling for which a significance of
3$\sigma$ is achievable.  It shows 3$\sigma$ sensitivity is achieved for couplings as low as  $\lambda'_{231}=0.3$ when conditions are most favourable.  We truncate the plot at $\lambda'_{231}=1.5$ due to the perturbativity limit described earlier.

\section{Conclusion}
\label{sec:conclusion}

Differences between $\mu^-e^+$ and $\mu^+e^-$ distributions have apparently received no attention at the LHC, even though they can potentially provide strong (greater than 10$\sigma$!)~evidence for BSM lepton flavour violation using only data-to-data comparisons.

We have demonstrated the above for models within the framework or RPV-supersymmetry.  Those models benefit from the fact that they have a large bias towards $\mu^-e^+$ production at the
LHC, while Standard Model backgrounds are expected either to be
symmetric or to (marginally) prefer $\mu^+e^-$.
\footnote{In the interests of more efficient phraseology in later works, it might be helpful if models could be termed `emu positive' or `emu negative' according to the sign of the muon that they prefer. According to such a convention, our claim is that the Standard Model would be `emu positive' and our $\lambda'_{231}$ model `emu negative'. While this nomenclature conflicts with the sign induced in $f$ (i.e.~an emu positive model induces negative $f$ and {\em vice versa}) it seems
appropriate considering that large flightless birds exist within the Standard Model.}

There are presumably other BSM models that pull in the same direction as the one considered, and yet more that will pull the other way.
Those in this latter camp may still be discovered by data-to-data comparisons of $\mu^-e^+$ and $\mu^+e^-$ distributions, however these will require the degree of SM bias to be explicitly determined before an observed asymmetry can be interpreted as a discovery.\footnote{Quantitative estimates of the SM bias will be useful for models in {\em both} bias `directions' if the systematic uncertainty on that bias in the signal region can be made smaller than its absolute magnitude.  In such a
case the
increased sensitivity that is bought by `subtracting' a large SM bias will not be offset by a larger systematic uncertainty on the  magnitude of that bias.}

We note that while we worked within the framework of dilepton events, and so were concerned with relationships between the {\em two} expectations as shown in equation~(\ref{eq:prop}), it seems likely that similar arguments could be made for appropriately defined comparisons of the {\em four} expectations $\left< N(e^+) \right>$, $\left< N(e^-) \right>$, $\left< N(\mu^+) \right>$ and $\left< N(\mu^-) \right>$ appropriate for single lepton events.
Some evidence in support of this statement may be found in Appendix~\ref{sec:singlelepton}.

Nonetheless, it should be possible to dig for detector-driven signatures in quite different areas altogether, so we hope this is only one of many directions in which future work could lead.

\clearpage

\acknowledgments
The authors gratefully acknowledge
the United Kingdom's Science and Technology Facilities Research Council (STFC) and Peterhouse for financial support, and
Ben Allanach, Sarah Driver, Thomas Gillam, Matthew Lim and other members of the Cambridge Supersymmetry Working Group for useful discussions.

\section*{Appendices}

\appendix

\section{Sources of charge-flavour bias considered for dilepton events}
\label{sec:biases}

\newcommand\nobias{Type 1}
\newcommand\lowbias{Type 2}
\newcommand\badbias{Type 3}

This section lists and categorises the potential sources of bias in the analysis, roughly in order of decreasing significance.
Herein $\rho$ refers always to the ratio of expectations found in equation~(\ref{eq:prop}).
Biases and effects are divided into three types:
\begin{itemize}
    \item
        \nobias:
those that leave $\rho$ invariant
\item
   \lowbias:
those that cannot increase $\rho$, and
\item
    \badbias:
those that can enlarge $\rho$ but cannot take it above one if already below one.
\end{itemize}

\subsubsection*{$W^\pm$ charge asymmetry in $W$+jet events}

The initial state in a proton-proton collider has an excess of positive over negative charge and a corresponding excess of valence up-quarks over valence down-quarks.  This asymmetry leads to a flavour-independent excess of
$u\bar d \rightarrow W^+ \rightarrow (e^+ \nu_e  \mathrm{\ or \ } \mu^+\nu_\mu)$
events, with or without extra jets, over
$\bar u d \rightarrow W^- \rightarrow (e^- \bar \nu_e  \mathrm{\ or \ } \mu^-\bar \nu_\mu)$ events \cite{Aad:2011yna,Aad:2011dm}.
Proton-proton collisions therefore show a preference for positively charged leptons in single lepton final states.  Our analysis requires two OSDF leptons, however, so if $W$+jet events are to pass our selection, the jets in the event must somehow produce a lepton of the opposite charge and flavour to that coming from the $W$.

\subsubsection*{Fake leptons}

Electrons are far more likely to be fake (e.g. jets mis-reconstructed as electrons) than muons. However, regardless of flavour, fakes are not biased to any particular charge.
For charge-symmetric processes, fakes thus add an equal contribution to the numerator and denominator of the ratio $\rho$, allowing it to be brought closer to unity, but not further. This makes the effect \badbias.
\hide{For example: if $\rho(f)=
\frac {a+f} {b+f}$ represents a ratio $\rho=\frac a b$ that is polluted by a fake component $f$ that is the same top and bottom, then \begin{align}\frac{d}{d f} \rho(f)&= \frac {b-a}{(b+f)^2} \end{align} is positive if $\rho<1$ and negative if $\rho >1$.  This shows that a ratio below one can be made bigger by increasing $f$, but since $\lim_{f \rightarrow\infty} \rho(f)=1$  it cannot be made to exceed one. Likewise a ratio that starts above 1 cannot be made, by the addition of
fakes, to end up lower than 1.}

The leading fake contribution to the $e \mu$ signature comes from single lepton processes in which one additional lepton is faked.
Since single lepton processes have a charge asymmetry, the differing rates of faking for electrons and muons could result in a bias to the ratio $\rho$.

There are essentially two main ways in which jets may produce leptons. In the case of electrons the dominant source will be misidentification, in the case of muons the dominant source will be heavy flavour decays.
The latter can be well suppressed at LHC detectors by requiring sufficient isolation. The former is harder to suppress. Accordingly we claim that \begin{align}
    p^W_\mu\ll p^W_e,
    \label{eq:oijniuhbuf}
\end{align}
where $p^W_\mu$ is the probability that a $W$+jet event will pass our selection due to a jet generating an isolated muon, and $p^W_e$ is the probability that a similar event will pass as a result of a jet faking an electron.

Why is this important?
Suppose that ratio $\rho$, prior to the consideration of the $W$+jet background, takes the value $a/A$, for some $0\le a \le A$.  Call this initial ratio $\rho_0$, i.e.~$\rho_0=\frac a A$. In terms of these quantities, the ratio {\em after} consideration of the $W$+jet background, $\rho_1$, would be expected to take the form:
$$
\rho_1 = \frac{
    a + k(N(W^-) \epsilon_\mu P(e^+) + N(W^+)  \epsilon_e P(\mu^-))
}{
    A + k(N(W^-) \epsilon_e P(\mu^+) + N(W^+)  \epsilon_\mu P(e^-))
}
$$
where: $N(W^\pm)$ are the expected number of $W^\pm$+jet events potentially inside acceptance; $\epsilon_e$ and $\epsilon_\mu$ are charge-independent efficiencies for reconstructing an isolated lepton or the relevant flavour from a $W$; $P(e^+)$ is the probability that a $W$+jet event has a jet that ends up looking like a positron; $P(e^-)$, $P(\mu^+)$ and $P(\mu^-)$ are the analogous quantities for other charges and flavours; and $k$ is an positive constant that accounts for the normalisation
definition used for $a$ and $A$ and the branching fraction of a $W$ to any species of light lepton. Given that fake electrons are not expected to prefer one charge over the other, we can say that:
$$P(e^+)=P(e^-)=p^W_e/2.$$
For $P(\mu^+)$ and $P(\mu^-)$ we cannot be quite so specific because muons from hadronisation could retain a small bias from the (on average positive) charge of the quarks from which the hadrons were formed.  For this reason we make only the weaker claim that: $$P(\mu^\pm)= {\kappa_\pm}\  p^W_\mu/2$$
where $\kappa_\pm$ are positive constants near one satisfying \begin{align}\kappa_- \le \kappa_+.
\label{eq:kappas}
\end{align} Given these statements we now have  that:
\begin{align}
    \rho_1
    &= \frac{
    a + \frac 1 2 k(N(W^-) \epsilon_\mu p^W_e + N(W^+)  \epsilon_e \kappa_- p^W_\mu)
}{
    A + \frac 1 2 k(N(W^-) \epsilon_e \kappa_+ p^W_\mu + N(W^+)  \epsilon_\mu p^W_e)
} \nonumber
\end{align}
Defining $N(\Delta^W)\equiv N(W^+)-N(W^-)$ and $k'\equiv k N(W^-)/2$ we can say further that
\begin{align}
    \rho_1
    &= \frac{
    a + k'( \epsilon_\mu p^W_e
    +
    \epsilon_e \kappa_- p^W_\mu
    +
    \frac{N(\Delta^W)}{N(W^-)}  \epsilon_e \kappa_- p^W_\mu
    )
}{
    A +
     k'(
      \epsilon_\mu p^W_e
    +
     \epsilon_e \kappa_+ p^W_\mu
    +
\frac{N(\Delta^W)}{N(W^-)}  \epsilon_\mu p^W_e \phantom{\kappa_-}\!\!)
} \nonumber
\end{align}
which can be written more succinctly as
\begin{align}
\rho_1 &= \frac {a + x + y+z}{A+X+Y+Z} \nonumber
\end{align}
if one defines
\begin{align} \nonumber
    x &=  k'  \epsilon_\mu p^W_e , \\ \nonumber
    X &=  k'  \epsilon_\mu p^W_e  , \\ \nonumber
    y &=  k'   \epsilon_e \kappa_-   p^W_\mu , \\ \nonumber
    Y &=  k'   \epsilon_e \kappa_+   p^W_\mu ,\\ \nonumber
    z &=  k' \frac{N(\Delta^W)}{N(W^-)}  \epsilon_e \kappa_-   p^W_\mu,\qquad\qquad\qquad\text{and} \\ \nonumber
    Z &=  k' \frac{N(\Delta^W)}{N(W^-)}  \epsilon_\mu    p^W_e .
\end{align}
We are now in a position to note that:
\begin{align} \nonumber
    \frac x X &=  1,  \\ \nonumber
    \frac y Y &=  \frac {\kappa_-} {\kappa_+} \le 1,\qquad\qquad\qquad\text{by (\ref{eq:kappas}),\phantom{x}and}    \\ \nonumber
    \frac z Z &= \frac {\epsilon_e \kappa_-   p^W_\mu} {\epsilon_\mu    p^W_e}\ll 1,
\end{align}
wherein the last step we have used both (\ref{eq:oijniuhbuf}) and the fact that $\epsilon_e$, $\kappa_-$  and $\epsilon_\mu$ are all numbers of order 1.
Since: (i) all of
$\frac x X$,
$\frac y Y$,
$\frac z Z$ and
$\frac a A$ have been shown to be less than or equal to one, (ii) at least one of them is less than one, and (iii) $a,x,y,z,A,X,Y$ and $Z$ are all positive, it is then trivial to show that
$\rho_1 = \frac{a+x+y+z}{A+X+Y+Z}<1$ and $\frac{x+y+z}{X+Y+Z}<1$.
The latter result proves that the bias from $W$+jet events is of \lowbias.

\subsubsection*{Other things related to the charge asymmetry of the $p$-$p$ initial state}

Above we showed that the $W^\pm$ asymmetry expected from the proton charge induces one of the desired forms of charge-flavour bias in dilepton events, even though it is `nominally' a mono-lepton background.  One should also consider the effect of the $pp$ initial state asymmetry in backgrounds containing $W$-bosons in which the secondary lepton is not fake or from heavy flavour, but is real. Backgrounds of this type, such as $W$+top, have biases that are much easier to categorise since all flavours are real and have predictable rates given by tree-level Feynman diagrams and universal weak lepton couplings. Most of these are therefore of \nobias.

\subsubsection*{Effects of detector geometry}
Detector geometry may induce acceptance differences depending on lepton charge.
As an example, the ATLAS muon system has a fixed toroidal magnetic field~\cite{AtlasMuonTdr}, 
with orientation such that the trajectories of $\mu^+$ and $\mu^-$ are bent oppositely in rapidity.
The authors could find no mention from the LHC collaborations of charge-dependence in the efficiencies for reconstruction of leptons, but this bending asymmetry could, in principle, lead to differences in acceptance or momentum resolution for positive and negative muons
in some parts of the detector\footnote{This might be expected to occur near transition regions in the detector or at the ends, where there are natural edges or changes in acceptance.}, as tracks differing only in charge may fall in regions of differing efficiency or may leave detector acceptance.
The effect is reversed in opposite ends of the detector, and so in a symmetric detector the bias disappears for event selections that are invariant under $\eta \leftrightarrow -\eta$.
There are effects which may disrupt this symmetry, however, for example a displacement of the interaction point from the geometrical centre of the detector, or an asymmetry in the active regions of the detector, either by design or by malfunction of detector components.

The magnitude of these effects is expected to be small for a number of reasons. Event selections are typically designed such that the edges of acceptance are avoided, giving relatively uniform efficiency~\cite{Aad:2016jkr}. In the case of regions of reduced efficiency, while muons of one charge may be lost at one edge of the anomaly, the opposite charge is lost at the other edge, largely nullifying the effect on the overall ratio.
Considering the position of the interaction point, while the LHC beam-spot is of finite size~\cite{AtlasBeamSpotResults}, this is expected to have little effect on the asymmetry when averaging over many interactions. Displacements of the beam-spot from the centre of the detector may be significant, but are typically small compared to the scale of the detector.
An attempt to estimate the magnitude of these effects will be made in Appendix~\ref{sec:quantestimates}.

\subsubsection*{Composition of matter}

Detectors contain electrons, but not positrons or muons (of either charge) and are therefore themselves charge-flavour asymmetric.
Charged particles traversing a detector can therefore kick out electrons (known as $\delta$-rays).  These are not expected to be reconstructed
as tracks in their own right \cite{ATLAS-CONF-2013-005}, however if they were they would present a source of charge-flavour bias that would appear on the denominator of $\rho$. Such biases are therefore at worst of \lowbias.

\subsubsection*{Charged pion decay}

It is well known that charged pions preferentially decay to $\mu \nu_\mu$ rather than to $e \nu_e$, even though the Standard Model's $W$-lepton-neutrino vertex is flavour-independent.\footnote{Angular momentum conservation and the handedness of the weak interaction, combined with the smallness of the electron mass compared to the muon mass, suppresses the decay to electrons more than to muons.} This effect is sometimes very important: muon neutrinos outnumber electron neutrinos in cosmic rays almost two to one because of it.
However, the effect is not expected to produce a significant excess of high energy isolated muons over  electrons in our search as it operates after hadronisation, meaning such muons would be soft and/or in jets. Were this effect nonetheless visible, it is in any case charge-symmetric and so should be of \nobias.

\subsubsection*{Cosmic ray composition and variation with depth}

For a number of different reasons, an excess of $\mu^+$ to $\mu^-$ is expected and observed in the underground muon flux from cosmic rays. The size of this ratio
increases with depth, and decreases with momentum, but is always positive
\cite{Adamson:2010xf,Adamson:2016vma}. This background is therefore presents a potential source of \lowbias\ bias.

\subsubsection*{Bulk shielding and beam backgrounds}
Material in and around detectors shields them from beam induced backgrounds.  These shields let muons pass more easily than electrons, but in a manner that is charge-independent, up to effects at the level of those that favour transmission of $\mu^+$ over $\mu^-$ in cosmic rays (see last paragraph). This muon over electron excess could additionally (marginally) favour positively-charged muons as another consequence of charged pions decays and the $W^\pm$ asymmetry already mentioned, and
so this source is \lowbias\ or \badbias.


\subsubsection*{$\frac {dE} {dx}$ differences between $e^+$ and $e^-$}

The rate of energy loss $\frac{dE}{dx}$ in matter is ever so slightly smaller for positrons than for electrons at relativistic energies (see equations (33.24) and (33.25) of \cite{Olive:2016xmw}).
This could mean that electron showers in calorimeters are, on average, slightly shorter than positron showers of the same energy. This effect could, in principle, create a small bias favouring containment of electron showers over those of positrons.  Put another way, at very high energies it is possible that the electron reconstruction efficiency could be a little higher than for positrons.  This potential bias, were it to exist, is the right way round to make it one of
\lowbias.


Though we list this as a {\em potential} source of bias, the experimental literature indicates that differences between electron and positron reconstruction efficiencies are at present unobservably small. See, for example, Figure 20 and associated text of \cite{Aad:2014fxa}.

\subsubsection*{$\frac {dE} {dx}$ differences between $\mu^+$ and $\mu^-$}

The rate of energy loss $\frac{dE}{dx}$ in matter is identical for $\mu^+$ and $\mu^-$ over all energies above 10~MeV (see Figure 33.1 of \cite{Olive:2016xmw}), and so this is not a source of bias for our study.

Relatedly, scale factors for tuning Monte Carlo predictions to match observations of data in control regions have likewise been found to be independent of charge where attempts to measure them have been made.  For example \cite{Aad:2011yna} reports:
\begin{quote}
    ... [these scale factors] are based on the ratio of the efficiency in data and in simulation, and are computed as a function of the muon $\eta_\mu$ and charge. The corrections for each charge agree within the statistical uncertainties, so the charge-averaged result is applied.
\end{quote}

Potential muon $\frac{dE}{dx}$ biases therefore seem to be non-biases, or equivalently are of \nobias.

\section{Quantitative estimates of biases}
\label{sec:quantestimates}

In this section, we shall give quantitative estimates of the leading sources of bias, as discussed qualitatively in the previous appendix.

\subsubsection*{\wfake\ charge-flavour asymmetry}

The magnitude of the bias introduced by fake leptons is dependent on the detector and identification algorithms applied. Experimental collaborations typically estimate these effects using data-driven methods, which would presumably be applied in an implementation of this analysis.
Here, we estimate the bias using a sample of simulated \wjets\ events in which the $W$ produces a muon, and a jet is treated as an electron with a chance of 0.5\% (similar to the rate reported by ATLAS in Ref.~\cite{ATLAS-CONF-2016-024}). This misidentification is assumed to be independent of the charge of the electron produced. The faking of muons is assumed to be negligible, making this an upper limit on the magnitude of the bias.

The contribution from \wfake\ events is substantial for low \summt, as can be seen in Figure~\ref{fig:osdfleptonpair:mthists}, but is reduced for the higher values sensitive to the $\lambda'_{231}$ signal models.
Above 250~GeV in \summt, \wfake\ makes up 5.5\% of the background in the \emmp\ channel, and 4.0\% in \epmm.
In the absence of the \wfake\ background, the ratio $\rho$ for the simulated background processes is consistent with unity.
Adding the \wfake\ background as estimated, this is lowered by 1.6\% for events with $\summt > 250~\text{GeV}$.

\subsubsection*{Other effects of $pp$ charge asymmetry}
Backgrounds producing two real leptons may be simulated by Monte Carlo. Those relevant are shown in Figure~\ref{fig:osdfleptonpair:mthists}, and, in the absence of the $W$+fake background discussed above, give a ratio $\rho$ consistent with unity.

\subsubsection*{Effects of detector geometry}
It was mentioned in Appendix~\ref{sec:biases} that asymmetries in the detector may induce a bias in $e^\pm\mu^\mp$ events.
Here we take the example of the ATLAS muon system~\cite{AtlasMuonTdr}, the fixed toroidal magnetic field of which bends the trajectories of $\mu^+$ and $\mu^-$ oppositely in rapidity.
The reconstruction efficiency as reported by ATLAS~\cite{Aad:2016jkr} is consistent to within 2\% over most of the $\eta-\phi$ plane, with the exception of the region $\lvert\eta\rvert<0.1$, where cabling and services enter the inner parts of the detector. We shall therefore focus our attention on this `gap', where efficiency is significantly lower, and examine the behaviour of a muon of relatively low transverse momentum (and so the track greatest curvature), $p_T=20~\text{GeV}$.
In this region, the toroid magnet has a bending power $\int B \cdot \text{d}l$ of approximately 3~Tm~\cite{AtlasMuonTdr}. We assume the field strength to be uniform within the toroid, giving a separation between positive and negative charges of $1.3^\circ$ ($\Delta\eta \sim 0.02$) on reaching the outer edge of the muon system. Given the inaccuracy in this modelling of the field, this charge separation has been enlarged to $\Delta\eta \sim 0.05$ in what follows.

In order to estimate the possible effect of detector anomalies, we consider a straw model in which muons of one charge within $\Delta\eta / 2$ of the $\lvert\eta\rvert<0.1$ gap are taken to be lost, while muons of the other charge are successfully reconstructed. This results in a change in the ratio $\rho$ over the whole detector of 0.7\%.
While illustrative, this number is pessimistic in several ways. Most notably, we have taken muons of only one charge to be lost. Under most circumstances, muons of the opposite charge will be lost at the other edge of a detector anomaly, reducing the effect on the overall ratio.

Another effect which may break the $\eta \leftrightarrow -\eta$ symmetry of the detector comes from possible displacements of the interaction point away from the geometrical centre of the detector.
While formerly $\eta$-symmetrical regions of reduced efficiency such as the $\lvert\eta\rvert<0.1$ gap lost as many muons of one charge as the other, a shift of the interaction point disrupts this.
Here we consider the case where the interaction centre is shifted by 45~mm along the $z$-axis, this being the typical radius of the LHC beam-spot as measured by ATLAS in 2015~\cite{AtlasBeamSpotResults}, and larger than the shift observed in the beam spot centroid for any five-minute period that year. This displacement corresponds to a shift in $\eta$ relative to the interaction point of roughly 0.01 for points close to $\eta=0$. Assuming that muons of one charge are lost within $\Delta\eta / 2$ of one edge, and the opposite charge at the other edge, very little asymmetry is induced. The change to the overall ratio $\rho$ is less than 0.1\%.

\section{Single lepton events}
\label{sec:singlelepton}

\begin{figure*}[tbp]
	\begin{center}
	\begin{subfigure}{.48\textwidth}
		\includegraphics[width=\textwidth]{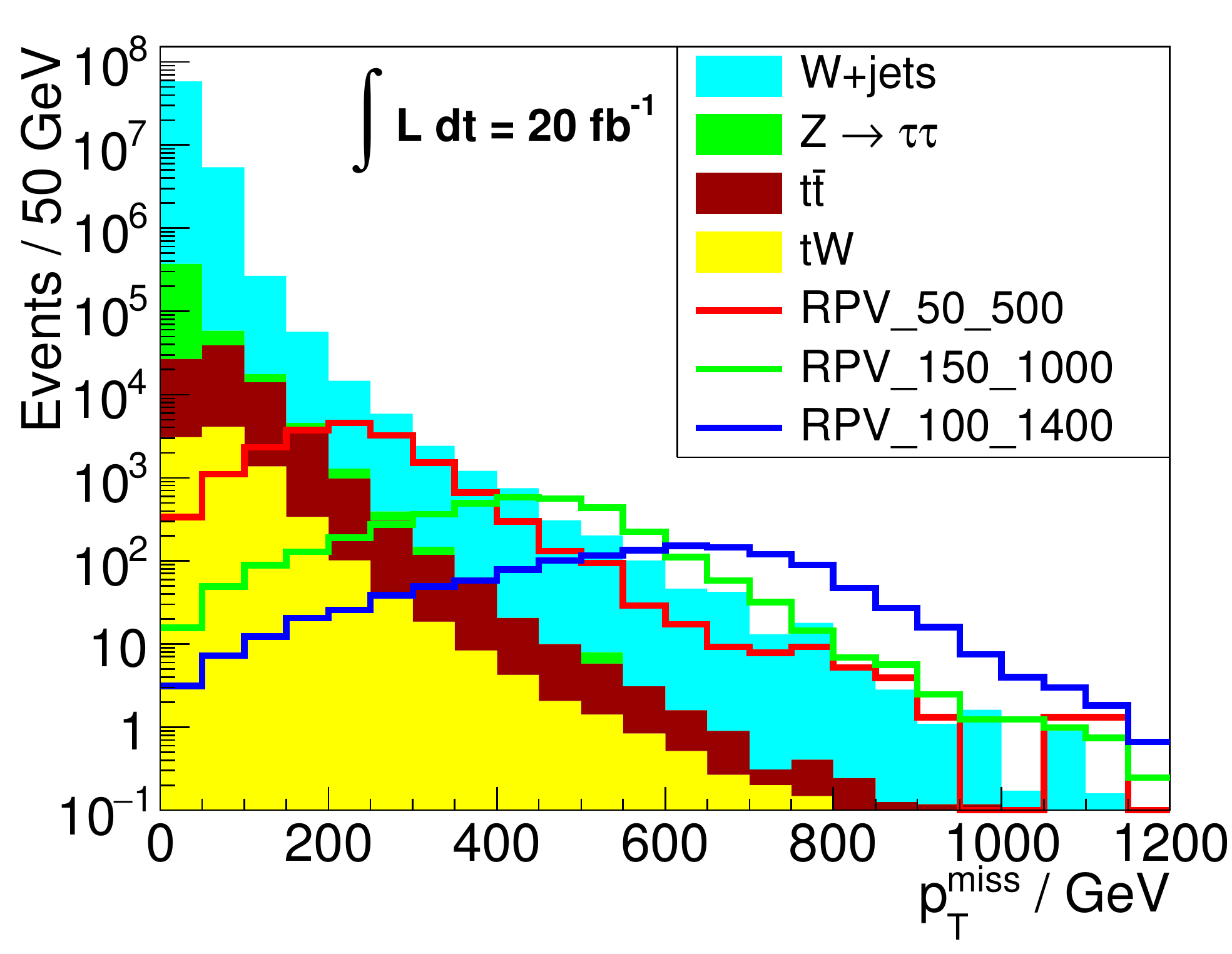}
			\caption{\ptmiss\ in $\mu^-$ events}
		\label{fig:singlelepton:methists:mum_met}
	\end{subfigure}
	\begin{subfigure}{.48\textwidth}
		\includegraphics[width=\textwidth]{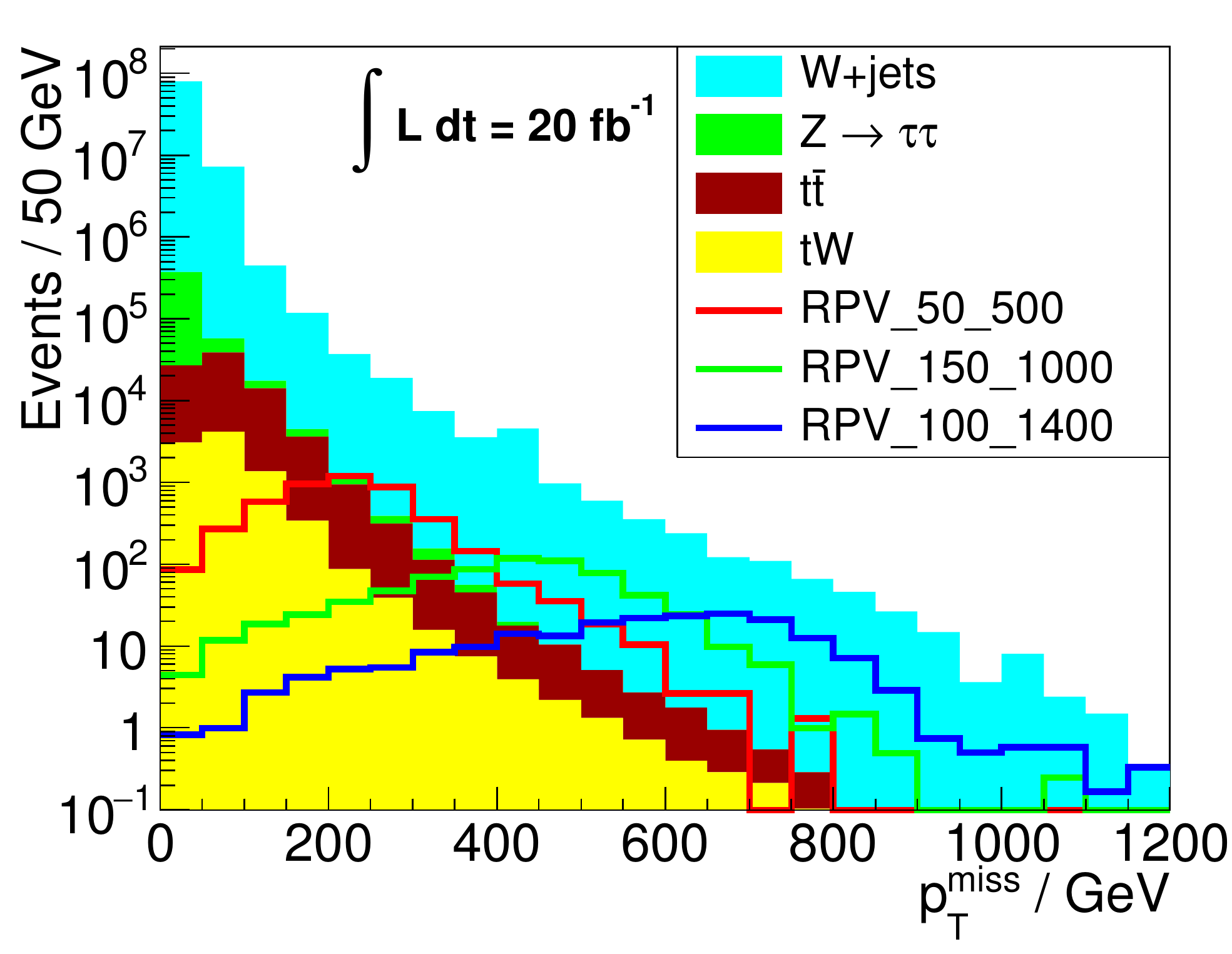}
			\caption{\ptmiss\ in $\mu^+$ events}
		\label{fig:singlelepton:methists:mup_met}
	\end{subfigure}
	\begin{subfigure}{.48\textwidth}
		\includegraphics[width=\textwidth]{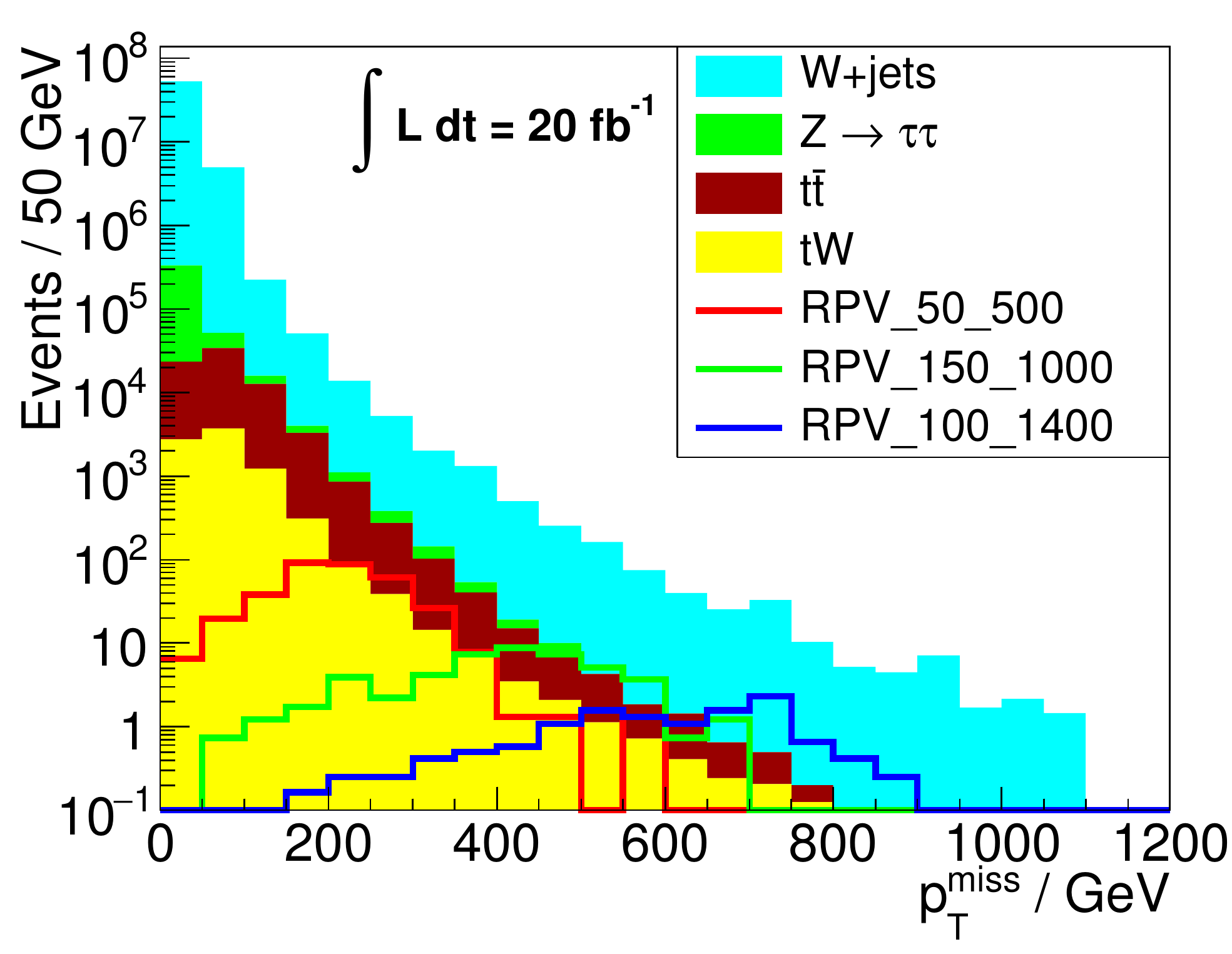}
			\caption{\ptmiss\ in $e^-$ events}
		\label{fig:singlelepton:methists:elm_met}
	\end{subfigure}
	\begin{subfigure}{.48\textwidth}
		\includegraphics[width=\textwidth]{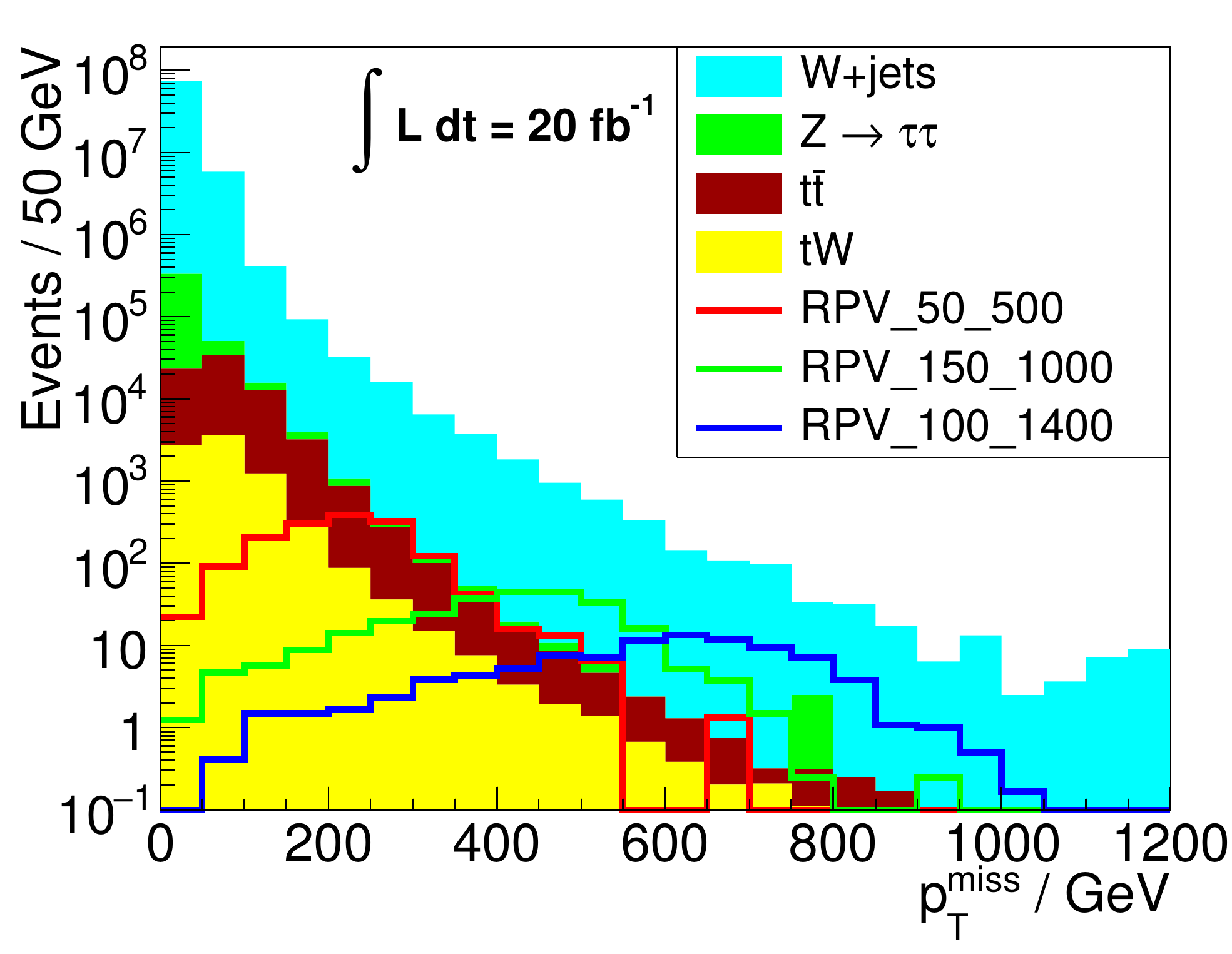}
			\caption{\ptmiss\ in $e^+$ events}
		\label{fig:singlelepton:methists:elp_met}
	\end{subfigure}
	\caption{ The distribution of \ptmiss\ in events in which the leading lepton is of a specific flavour and charge.
	}
	\label{fig:singlelepton:methists}
	\end{center}
\end{figure*}

It might be possible to exploit charge-flavour asymmetries in single lepton events instead of (or in addition to) the dilepton events considered in the rest of the paper.

Figure~\ref{fig:singlelepton:methists} shows \ptmiss\ distributions for electron and muon events of each charge separately, for the usual three signal models and a variety of backgrounds.
Here the dominant background contribution shown comes from Standard   Model production of $W$~bosons, which is simulated in slices of \ptmiss.
Top-pair and $tW$ processes and are also included.
This figure shows that the signal samples have a significant dependence on lepton charge and flavour, and that $\lambda'_{231}$ induces a much larger cross section for negatively-charged muons than for any other flavour or charge of lepton. This is to be contrasted with what appears to be (relatively) a much smaller dependence on charges and flavour in the SM backgrounds.

On account of the proton-proton-induced $W^\pm$ charge asymmetry the positively- and negatively-charged SM backgrounds are expected to (and do) differ, but such differences
are themselves expected to be flavour symmetric and so are ripe for cancellation under an appropriate modification of the notions of weak and strong `conspiracy' for single lepton events.

Nonetheless, these plots should be regarded as little more than a source of encouragement to investigate further; single lepton events have a significant background that is not on these plots, and which is more important for single leptons than it is for dileptons.  These are fake leptons.  While fakes should be charge symmetric, they are tricky to simulate, potentially large in number, and could be associated with broad tails in \ptmiss. We elect to leave the question of whether charge-flavour differences are observable in single-lepton events
(after inclusion of all other relevant backgrounds and consideration of trigger issues)
as a topic for future study.

\bibliographystyle{JHEP}
\bibliography{paper}

\end{document}